\documentclass[article]{aa}  

\usepackage{graphicx}
\usepackage{txfonts}
\usepackage{lscape}
\usepackage{booktabs}
\usepackage{threeparttablex}
\usepackage{wasysym}
\usepackage{float}
\usepackage{diagbox}
\usepackage{capt-of}
\usepackage{caption}
\usepackage{amssymb}
\usepackage{arydshln}
\usepackage{natbib}
\usepackage{rotating}
\usepackage{xcolor}
\usepackage[bookmarks=false,colorlinks=true,linkcolor=cyan,citecolor=blue,filecolor=black,urlcolor=cyan]{hyperref}
\newcommand*{\sfr}{\ensuremath{\mathrm{SFR}}}
\newcommand*{\stellarmass}{\ensuremath{\mathrm{M}_{\star}}}
\newcommand*{\lx}{\ensuremath{\rm L_{X}}}
\newcommand*{\solarmass}{\ensuremath{\mathrm{M}_{\odot}}}

\newcommand*{\luminosityunits}{\ensuremath{\mathrm{erg\,s^{-1}}}}
\newcommand*{\energyunits}{\ensuremath{\mathrm{keV}}}
\newcommand*{\distanceunits}{\ensuremath{\mathrm{Mpc}}}

\begin{document} 

   \title{Prescriptions for the stochasticity effect on the integrated X-ray luminosity of star-forming galaxies}
   
   \subtitle{Implications for selecting star-forming galaxies and AGN in X-ray surveys}
   
   \author{E. Kyritsis \inst{1,2,3}
          \and
          A. Zezas\inst{1,2}
          \and
          K. Kovlakas\inst{4,5}
          }
          \authorrunning{Kyritsis et al.}

   \institute{Physics Department, \& Institute of Theoretical and Computational Physics, University of Crete, GR 71003, Heraklion, Greece\\
   \email{ekyritsis@physics.uoc.gr}
   \and
    Institute of Astrophysics, Foundation for Research and Technology-Hellas, GR 71110 Heraklion, Greece
    \and
    Max-Planck-Institut für Extraterrestrische Physik, Gießenbachstraße 1, 85748 Garching, Germany
    \and
    Institute of Space Sciences (ICE, CSIC), Campus UAB, Carrer de Magrans, 08193 Barcelona, Spain
    \and
Institut d’Estudis Espacials de Catalunya (IEEC), Edifici RDIT, Campus UPC, 08860 Castelldefels (Barcelona), Spain
    }

   \date{Received Month XX, XXXX; accepted Month XX, XXXX}

% \abstract{}{}{}{}{} 
% 5 {} token are mandatory
 
  \abstract
  % context heading (optional)
  % {} leave it empty if necessary  
   {The integrated X-ray luminosity (\lx{}) of star-forming galaxies originates primarily by populations of high-mass X-ray binaries (HMXB). However, the discrete nature of these populations introduces stochastic sampling effects, affecting the shape of their underlying X-ray Luminosity function (XLF), and biasing the interpretation of observed scaling relations.}
  % aims heading (mandatory)
   {In this work we investigate how stochastic sampling of the HMXB XLF impacts the predicted integrated \lx{} of galaxies across a broad range of star-formation and metallicity conditions, and quantify the resulting scatter in order to provide a statistical framework for interpreting X-ray observations of galaxies}
   {By performing Monte Carlo (MC) simulations  we derived distributions of the integrated \lx{} of galaxies over a wide grid of star-formation rate (\sfr{}) and metallicity, covering the broad range of conditions observed in star-forming galaxies. By measuring statistical quantities that delineate the complex shapes of these distributions, we parametrized the luminosity scatter by fitting surfaces to the upper and lower \lx{} bounds as function of \sfr{} and gas-phase metallicity.}
   {We provide a practical set of prescriptions to calculate the expected integrated \lx{} for a given \sfr{} and metallicity, fully accounting for stochastic effects without the need to rerun the computationally expensive sampling of the HMXB XLF for each individual galaxy. Application of these prescriptions to local and higher-redshift galaxy samples, shows that stochasticity must be considered before attributing differences in \lx{} to intrinsic population properties. Furthermore, a galaxy simulation study across z =0.5–5 showed mild intrinsic redshift evolution of the stochastic scatter, mirroring the evolution of \sfr{} and metallicity, with minimum scatter at z$\sim$2.5 where the cosmic star-formation density reached its maximum. In addition, our prescritions can be used to quantify the bias introduced in the redhsift dependent \lx{}-\sfr{}-Metallicity scaling relations due to the flux-limited nature of the X-ray surveys.  
   Finally, we find that at low redshifts, stochastic effects can raise \lx{} by up to 1 dex with respect to the mean relation, leading to overlap with the regime of low-luminosity AGN (LLAGN), thus potentially biasing the classification of X-ray sources in deep-field surveys.}
  % conclusions heading (optional), leave it empty if necessary 
   {Our results demonstrate that stochastic sampling of the HMXB XLF is a fundamental source of scatter in the \lx{}/\sfr{} relation, especially at low \sfr{}s. The intrinsic scatter shows a mild evolution with redshift, and its imprint is modulated by the evolving \sfr{} and metallicity of galaxies across cosmic time. Accounting for these effects is essential for interpreting the X-ray emission of galaxies, exploring potential biases of the current scaling relation imposed by the flux-limited surveys, and disentangling X-ray emission from normal star-forming galaxies from AGN activity in deep X-ray surveys. The prescriptions presented here offer a practical framework for constraining the scatter around the scaling relations, quantifying the likelihood of extreme outliers, and refining the classification of X-ray sources in current and future surveys.}
   \maketitle

%-------------------------------------------------------------------
\section{Introduction}\label{sec:intro}
The total X-ray emission from normal galaxies (i.e. those not harbouring active galactic nuclei-AGN), originates from discrete sources, such as X-ray binaries (XRBs), and hot diffuse gas. In galaxies with high star-formation activity, the hard X-ray emission ($>$2 \energyunits) is pridominantly driven by populations of high-mass X-ray binaries (i.e HMXBs), which consist of a compact object (a black hole or a neutron star) accreting material from a high-mass companion star \citep[M$>$8 \solarmass{};][]{reig11}. In lower energies ($<$2 \energyunits) the relative contribution varies depending on the \sfr{} and the \stellarmass{} of the galaxy.
As this X-ray emission traces the underlying stellar populations, it provides a powerful tool for studying the endpoints of stellar evolution and the formation and evolution of, the otherwise invisible, compact object populations \citep[see][and references therein]{gilfanov22,zezas26}. Given that HMXBs can be a phase in the evolution of the progenitors of gravitational wave sources and short gamma-ray bursts \citep[sGRBs;][]{marchant16}, studying their X-ray emission offers key insights into the physical mechanisms that govern these phenomena and enables predictions of their formation rates. Furthermore, the cumulative X-ray output from HMXBs plays a significant role on their host galaxy and its surrounding intergalactic medium. This is particularly important in the early Universe, where the high-energy photons emitted by HMXBs in dwarf galaxies, due to their long mean free paths, may have contributed significantly to the radiation responsible for preheating of the intergalactic medium and the formation of the first galaxies during the epoch of reionization \citep{das17, madau17,eide18,kovlakas22}.

A powerful tool for probing the statistical properties of HMXB populations is through the construction of their X-ray luminosity  functions (XLFs). Over the past two decades, deep and high-resolution observations with Chandra and XMM-Newton have enabled significant progress in this direction, allowing the detection of individual HMXBs in nearby galaxies and the characterization of their luminosity distributions across different galactic environments. These studies have shown that both the shape and normalization of the HMXB XLF correlate strongly with the properties of the host galaxy, such as the star formation rate (SFR) and stellar mass (\stellarmass{}) \citep{grimm02,gilfanov04,mineo12a,zhang12,lehmer19,lehmer20}. More recent observational and theoretical works show that metallicity and star formation history also affect the XLF \citep{basuzych13b, fragos13a, fragos13b, kovlakas20, lehmer21, gilbertson22,misra23, geda24}.

Understanding the XLF and its dependence on stellar population parameters of the host galaxy is essential. However, while the study of individual sources is a powerful tool, in many cases, and especially in more distant galaxies, it is not possible to resolve individual binaries or reach luminosities that yield statistically useful samples. In such cases, the integrated emission of XRBs in galaxies provides an alternative powerful approach for studying their connection with stellar populations. Specifically, it provides a direct link between the galaxy’s total X-ray luminosity (\lx{}) and its stellar populations, resulting in the empirical scaling relations, such as the $L_{\mathrm{X}}$–\sfr{}–\stellarmass{}–metallicity relation for star-forming galaxies. 

These relations are widely used to trace star formation and binary evolution across cosmic time \citep{mineo14, lehmer16, kouroumpatzakis20, riccio23,kyritsis25}, as well as, to dissentangle emission from XRBs or accreting supermassive black holes \citep[SMBHs;][]{mezcua18, latimer21,sacchi24}. However, the total \lx{} observed in any given galaxy represents a stochastic realisation of a finite number of HMXBs sources drawn from the underlying XLF. This inherent statistical nature of the integrated XLF introduces intrinsic scatter around the average scaling relations and can lead to significant biases of the scaling relations, particulalry at the low-\sfr{} and low-\stellarmass{} regime.

Characterising and quantifying this scatter is especially important for low-SFR galaxies, where only a few HMXBs are expected and as such, their integrated X-ray emission is dominated by stochastic effects \citep{gilfanov04,anastasopoulou19,lehmer19,kyritsis25}. In such cases, estimating an upper limit on the total expected X-ray luminosity, accounting for both stochasticity and host galaxy properties (\sfr{}, metallicity), helps determine whether the observed emission is consistent with an assumed XLF or indicative of rare, luminous sources that may suggest differences in the shape of the XLF. For instance, X-ray bright dwarf galaxies may host ultraluminous X-ray sources (ULXs) accreting at super-Eddington rates \citep{kaaret17}, offering insights into extreme accretion regimes, and the demographics of compact objects at the upper end of the mass spectrum.

On the other hand, setting robust upper bounds (i.e. accounting for the stochastic scatter) on the expected $L_{\mathrm{X}}$ from XRB populations can help to constrain the identification of low-luminosity AGN (LLAGN) which may lack optical or infrared signatures. These systems often exhibit X-ray luminosities comparable to those of extremely luminous XRB populations found in dwarf metal poor starburst galaxies \citep{fornasini18,kouroumpatzakis21,kyritsis25}. Accurately distinguishing between HMXBs and LLAGN is particularly important for deep and wide area X-ray surveys, where source classification often relies on a fiducial luminosity threshold (e.g. $L_{\mathrm{X}} > 10^{41}$ - $10^{42}$  erg s$^{-1}$). In addition, since this scatter is the result of the stochastic sampling of the XLF and depends on SFR and metallicity, both of which evolve with redshift \citep{madau14}, it is important to investigate whether the scatter itself evolves over cosmic time. This is particularly relevant for deep surveys, where detected galaxies may represent positive statistical fluctuations near the sensitivity limit, and hence may bias the measured scaling relations between X-ray luminosity and stellar populations. In addition, it may affect the power spectrum of the 21 cm signal from the epoch of reionization \citep{kovlakas22,nikolic24}.

In order to understand the stochastic effects on the total \lx{} introduced by the incomplete sampling of the HMXB XLF, one can adopt the statistical framework presented in \cite{gilfanov04} which describes the behaviour of the integrated luminosities in systems dominated by discrete sources. Given a differential HMXB-XLF $\rm \frac{dN_{\rm{HMXB}}}{dL}$, the expected number of HMXBs in a galaxy above a minimum luminosity ($L_{
\rm{min}}$) is:
\begin{equation}\label{eq:Nexp}
\rm N_{\rm{exp}} = \int^{L_{\rm{max}}}_{L_{\rm{min}}} \frac{dN_{\rm{HMXB}}}{dL} dL \ ,
\end{equation}
and the corresponding expected total luminosity is given by:
\begin{equation}\label{eq:Ltotexp}
\rm L_{\rm{X,tot}} = \int_{L_{\rm{min}}}^{L_{\rm{max}}} L \frac{dN_{\rm{HMXB}}}{dL} dL 
\end{equation}
where $\rm L_{\mathrm{max}}$ denotes the upper luminosity which ensures the convergence of the HMXB XLF integration.

Assuming that the differential HMXB-XLF follows a power-law  with an index $\alpha$ and it is proportional to the \sfr{}, that is $\rm \frac{dN_{\rm{HMXB}}}{dL} \propto \rm SFR \times L^{-\alpha} $, the expected total luminosity is:
\begin{equation}\label{eq:Ltotexp_prop_sfr}
\rm L_{\rm{X,tot}} = \int_{L_{\rm{min}}}^{L_{\rm{max}}} L \frac{dN_{\rm{HMXB}}}{dL} dL \propto \frac{\rm SFR}{2 - \alpha}\left( L_{\mathrm{max}}^{2 - \alpha} - L_{\mathrm{min}}^{2 - \alpha} \right). 
\end{equation}

According to the Eq.~\ref{eq:Ltotexp_prop_sfr}, when the slope of the XLF lies in the range $1 < \alpha < 2$, as is the case for HMXBs \citep[$\alpha \sim 1.6$;][]{grimm03,mineo12a}, the total luminosity is not simply linearly proportional to the number of sources or, equivalently, to the \sfr{}. Instead, the integrated luminosity is dominated by the high-luminosity end of the XLF (i.e. the $\rm L_{\mathrm{max}}^{2 - \alpha}$ term), which drives the stochastic presence of a few bright, rare sources. 

This behaviour has two important implications. First, in low-SFR galaxies where only a small number of HMXBs are expected, the total luminosity becomes highly sensitive to stochastic fluctuations in the number and luminosity of the brightest sources. In some galaxies, one or two luminous HMXBs may dominate the total X-ray emission, while in galaxies where no high-luminosity sources are drawn from the XLF due to small-number statistics, the observed luminosity may fall well below the expected mean. Second, the distribution of the total luminosities for a given SFR is strongly skewed, exhibiting a long tail toward high luminosities. As a result, the mode and median of this distribution are systematically lower than the ensemble average. This effect introduces a super-linear trend in the $L_{\mathrm{X}}$–SFR relation at the low-SFR regime and contributes to the observed scatter around the scaling relations \citep{kouroumpatzakis20, geda24}. Understanding and quantifying the impact of stochastic sampling is thus crucial for robustly linking a galaxy’s X-ray luminosity to its stellar populations.

In recent years, several studies have explored the role of stochastic sampling on the integrated X-ray luminosity of star-forming galaxies. Using Chandra and ASCA observations of nearby starburst galaxies, \cite{grimm03} proposed a differential HMXB XLF which scales linearly with the \sfr{}, and it described by a power-law with index $\alpha \sim 1.6$ and an upper cutoff X-ray luminosity of a few $\times 10^{40}$ \luminosityunits. By integrating this XLF they showed that in low‑\sfr{} galaxies (i.e. \sfr{}$ \lesssim 4.5 ~\rm M_{\odot}\ yr^{-1}$), the total X-ray luminosity scales super-linearly with the \sfr{}. This deviation from linearity was attributed to stochastic fluctuations in the number and luminosity of HMXBs sampled from the underlying HMXB XLF, with a few bright sources dominating the integrated luminosity of these galaxies. On the other hand, in galaxies with higher \sfr{}s, the HMXBs population becomes large enough to average out the stochastic effects, resulting in a linear relation. 

\cite{lehmer19} refined these constraints on the HMXB-XLF using a sample of 38 nearby galaxies ($\rm D = 3-29$ \distanceunits) observed with Chandra. Based on their best-fitted global XLF model they performed Monte Carlo (MC) simulations of the expected X-ray luminosity distribution for a given \sfr{} to evaluate whether the observed \lx{} could be explained by stochastic sampling of the XLF. Their results showed that a population of low-metallicity galaxies, significantly deviates from their model beyond the statistical scatter, indicating that the XLF also depends on metallicity.

To that end, \cite{lehmer21} (hereafter L21) investigated explicitly the role of metallicity in shaping the HMXB-XLF. Using a sample of 55 nearby, actively star-forming galaxies observed with Chandra, they proposed a metallicity-dependent HMXB XLF, proportional to the \sfr{} which described by a broken power law with a metallicity-independent faint-end slope, a metallicity-dependent bright-end slope, and an exponential cut-off luminosity. To quantify the effects of stochastic sampling in low-SFR galaxies they followed a similar MC analysis as in \cite{lehmer19}, and they calculated the expected distribution of X-ray luminosity per unit SFR using their best metallicity-dependent HMXB XLF model. Their results (L21, Fig. 5) show that for galaxies with SFR $\lesssim 2~\mathrm{M_{\odot}~yr^{-1}}$, the predicted \lx{}/SFR distributions are broad and highly skewed, driven by stochastic variations in the number and luminosity of bright HMXBs. In contrast, at higher SFRs, the distributions become approximately Gaussian, centered around the mean expected from the integration of the XLF, with reduced scatter.

Most recently, a number of studies have followed similar approaches to quantify the impact of stochastic sampling on the integrated X-ray luminosity by performing MC simulations of the HMXB XLF on a galaxy-by-galaxy basis, using the observed \sfr{} and/or metallicity of each system \citep[i.e.][]{anastasopoulou19,geda24,kyritsis25}. For instance, \citet{geda24} applied the metallicity-dependent XLF model of L21 to a sample of nearby dwarf galaxies observed with Chandra (\stellarmass{}$< 5 \times 10^{9}~\mathrm{M_{\odot}}$, $\rm D < 12.5$~\distanceunits). These galaxies were selected to be among the most likely hosts of LLAGN with X-ray output similar to that of HMXBs. Their MC analysis confirmed that in the low-SFR regime, the predicted \lx{}/SFR distributions exhibit substantial stochastic variability, consistent with earlier studies. In addition, they showed that the probability of a galaxy hosting a very bright HMXB (one that could be confused with an LLAGN) is significantly reduced. 

However, as indicated by Eq.~\ref{eq:Nexp}, this approach requires a first level of MC simulations, by applying repeated random sampling from the HMXB XLF to determine the expected number of sources. Given that the normalization of the XLF depends on the \sfr{}, this procedure can become computationally intensive, with runtimes ranging from a few minutes per galaxy at low \sfr{} to several hours at high \sfr{}. As a result, applying this methodology to large samples or to galaxies with high \sfr{} can be prohibitively time-consuming. To overcome this limitation, this work provides, for the first time, a set of prescriptions that enable the estimation of the expected \lx{} at different confidence intervals  from stochastic sampling of the HMXB XLF, without the need to rerun the full suite of simulations for each individual galaxy. We achieve this by performing extensive MC simulations over a wide grid of \sfr{} and metallicity values. We then use the resulting X-ray luminosity distributions to parametrise the stochastic scatter as a function of \sfr{} and metallicity. 
%--------------------------------------------------------------------------------------------------------------
\section{New prescriptions for the stochastic scatter on the integrated X-ray luminosity of normal galaxies }\label{sec:sample_construction}

To parametrise the stochastic scatter as a function of the stellar population of the host galaxy, we performed a MC simulation study to compute the expected integrated X-ray luminosity distributions arising from stochastic sampling of the HMXB XLF over a grid of \sfr{} and metallicity values. Then by using characteristic statistical quantities derived from these distributions, we modeled the relation between the corresponding X-ray luminosity in different confidence intervals, the \sfr{} and the metallicity of star-forming galaxies. All the X-ray luminosities reported in this work are calculated in the  $0.5 - 8.0$ keV energy band, while metallicity refers to the gas-phase abundance, expressed as $[12 + \log(\mathrm{O/H})]$. In the following sections we present all the steps taken for this procedure. 

\subsection{Calculation of the expected X-ray luminosity distributions based on the HMXBs XLF}\label{subsec:lx_distributions}
As a first step, we constructed the \sfr{}-metallicity grid. Since stochastic effects are most pronounced in the low-\sfr{} regime, we adopted a non-uniform sampling strategy in which the grid has high resolution at low \sfr{}s and coarser resolution at higher \sfr{}s, covering the full range from \sfr{} $= 0.001$ to $100 ~\rm M_{\odot}~yr^{-1}$. For the gas-phase metallicity, we adopted a uniform sampling with a fixed step of 0.25 dex across the range $[12 + \log(\mathrm{O/H})] =$ 7.0-9.0 . The resulting \sfr{}–metallicity grid is presented in Table~\ref{tab:SFR-metal grid}. This procedure yields 144 \sfr{}–metallicity pairs, representative of those observed in star-forming galaxies. 

For each \sfr{}-metallicity pair, we performed a MC analysis by simulating the X-ray emission of 20\,000 fiducial galaxies. The simulation was performed in two steps. First, we calculated the expected number of HMXBs, $N_{\rm exp}$, by integrating the metallicity-dependent HMXB XLF of L21 using Eq.~\ref{eq:Nexp}. For the integration we used the \texttt{quad} function from the \texttt{scipy} python library \citep{virtanen20} and the integration limits were between $\rm L_{\rm{min}} = 10^{36}\,\rm{erg}\,\rm{s^{-1}}$ and $\rm L_{\rm{max}} = 5\times10^{41}\,\rm{erg}\,\rm{s^{-1}}$, corresponding to the luminosity range over which the metallicity-dependent HMXB XLF of L21 is defined. We note that changing the low integration limit had little effect on the final result. To account for model uncertainties in the XLF shape, we treated the key best-fit parameters, the normalization, the metallicity-independent slope, and the metallicity-dependent slope, as gaussian distributions\footnote{We note, that by ignoring the covariance between the L21 HMXB XLF parameters leads to a conservative approach, resulting in a overestimation of the intrinsic scatter.} with means equal to their best-fit values and standard deviations the 16\%-84\% confidence range as they are quoted in Table 2 of L21. All other parameters were fixed at their best-fit values.
In that way, we produced a list of 20\,000 N$_{\rm exp}^{\rm i}(\rm SFR,[12+log(O/H)]$ values per \sfr{}-metallicity pair.

As a second step, we accounted for Poisson fluctuations in the discrete number of sources, by drawing 20\,000 instances of the actual number of HMXBs as:
\begin{equation}
\rm N_{\rm exp,i}^{\rm inst} \sim \rm Pois(N_{exp}^{i}(SFR,[12 + \log(\mathrm{O/H})])),\quad\quad\text{i=1,...20\,000}.
\end{equation}
For each instance, we sampled the L21 HMXB XLF by drawing the \lx{} of $\rm N_{\rm exp,i}^{\rm inst}$ sources, and computed the total X-ray luminosity of the overall XRB population as 
% (accounting for its uncertainties)
\begin{equation}
\rm L_{\rm tot\,exp, i}^{\rm inst} = \sum_{j=1}^{{\rm N_{\rm exp, i}^{\rm inst}}}\rm L_{X}^{j},\quad\quad\text{i=1,...20\,000}
\end{equation}
where $\rm L_{\rm X}^{\rm j}$ is the luminosity of the $\rm j^{th}$ source drawn from the XLF. 

\begin{figure*}
 \centering
        \includegraphics[width=0.8\linewidth]{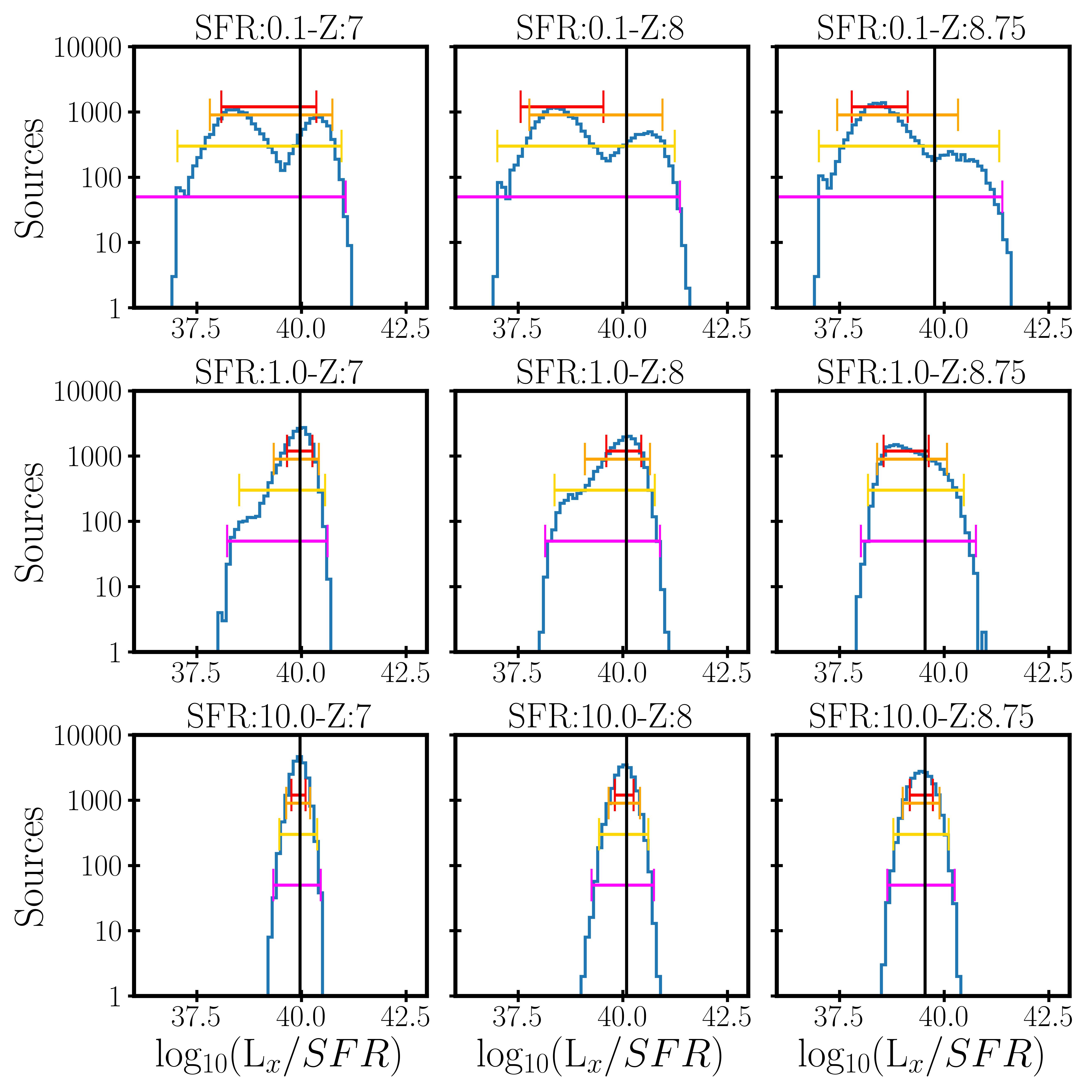}
    \caption{Indicative X-ray luminosity distributions for low, moderate, and high SFR-[12+log(O/H)] pairs in our grid. For better visualization, we denote in the plot the gas-phase metallicity [12+log(O/H)] as Z. The red, orange, yellow and purple error-bars indicate the $68\%$, $90\%$, $99\%$, and $99.9\%$ HDIs, respectively. The vertical solid black line shows the mean \lx{} of each distribution.} 
    \label{fig:samples_distribution}
\end{figure*}

Based on this procedure, we obtained 144 X-ray luminosity distributions, and each of them contains $20\,000$ integrated luminosities. 
These distributions correspond to every \sfr{}–metallicity pair in our grid, and incorporate both fluctuations in the number of sources per galaxy and stochastic effects from sampling the XLF. In Fig.~\ref{fig:samples_distribution} we present the simulated X-ray luminosity distributions per \sfr{} and metallicity. For each distribution, we measured the lower and upper bounds of the $68\%$, $90\%$, $99\%$, and $99.9\%$ highest density credible intervals (HDIs) using the Bayesian toolkit ArviZ \citep{arviz19}. These intervals provide a quantitative measure of the scatter and they are particularly well suited for complex, skewed, or bimodal distributions such as those in our analysis. In addition, we computed both the mean and the mode of each distribution. For each distribution, we also overplot the $68\%$, $90\%$, $99\%$, and $99.9\%$ HDIs and the mean values. 

\begin{table}
\centering
\caption{The \sfr{} - [12+log(O/H)] grid considered in this work.}
\begin{tabular}{cc}
\hline\hline
\sfr{} ($\rm M_{\odot}~yr^{-1}$) & [12+log$(\rm O/H)$]  \\
\hline
0.001 & 7 \\
0.005 & 7.25 \\
0.01 & 7.5 \\
0.025 & 7.75 \\
0.05 &  8 \\
0.75 &  8.25 \\
1 & 8.5 \\
2.5 & 8.75 \\
5 & 9 \\
10 & - \\
25 & - \\
50 & -\\
100 & - \\
\hline
\end{tabular}
\label{tab:SFR-metal grid}
\end{table}

\subsection{Stochasticity effects on the shape of the X-ray luminosity distributions of star-forming galaxies.}\label{subsec:x-ray_distributions}
In Fig.~\ref{fig:samples_distribution} we present indicative X-ray luminosity distributions resulting from stochastic sampling of the L21 HMXB XLF for typical low, moderate, and high SFR–metallicity pairs in our grid. The evolution of the shape of the distributions with \sfr{} and metallicity reflects the impact of stochasticity on the integrated luminosity of star-forming galaxies. More specifically, at moderate \sfr{} (i.e. \sfr{}$=1 \ \rm M_{\odot}\ yr^{-1}$), the distributions are strongly skewed, reflecting the dominant contribution of rare bright sources on the galaxy's X-ray output. In contrast, at high \sfr{}s (i.e. \sfr{}$=10\,\rm M_{\odot}\,yr^{-1}$) the distributions progressively approach a Gaussian-like form as expected from the central limit theorem given the larger number of contributing sources. In addition, in the very low-\sfr{} regime (i.e. \sfr{}$\leq 0.1\,\rm M_{\odot}\,yr^{-1}$) most realizations result in no sources (e.g. N$_{\rm exp}$ = 0) with \lx{}$>10^{37}\,\rm erg\,s^{-1}$ (not shown in Fig.~\ref{fig:samples_distribution}) highlighting the imprint of stochasticity at a galaxy-by-galaxy level. For instance, a galaxy may host a single bright HMXB dominating its luminosity, while another galaxy with identical properties (i.e. \sfr{}, [12+log(O/H)]) may host none.

Furthemore, we can see the effect of the metallicity on the shape of the integrated \lx{} distributions. For example, at fixed \sfr{}=1 $\rm M_{\odot}\,yr^{-1}$, the mean luminosity shows little variation, but the \lx{} changes significantly, from highly bimodal at low metallicity to strongly skewed at higher metallicity. This behaviour can be explained by the metallicity-dependence of the L21 HMXB XLF. In low-metallicity galaxies, the flatter high-luminosity end of the XLF enhances the probability of sampling more X-ray luminous HMXBs compared to metal-rich galaxies of the same \sfr{}. 

The differences in the shape of the \lx{} distributions are also reflected in the HDIs which evolve with the \sfr{} and metallicity. Specifically, low-\sfr{}s lead to HDIs which for the different confidence levels do not have the regular differences seen in Gaussian or other symmetric distributions. On the other hand, at high-\sfr{}s where the distributions are nearly Gaussian and the HDIs closely match their intrinsic scatter.

\begin{figure}
\includegraphics[width=0.5\textwidth, trim={5cm 2cm 0cm 0cm},clip]{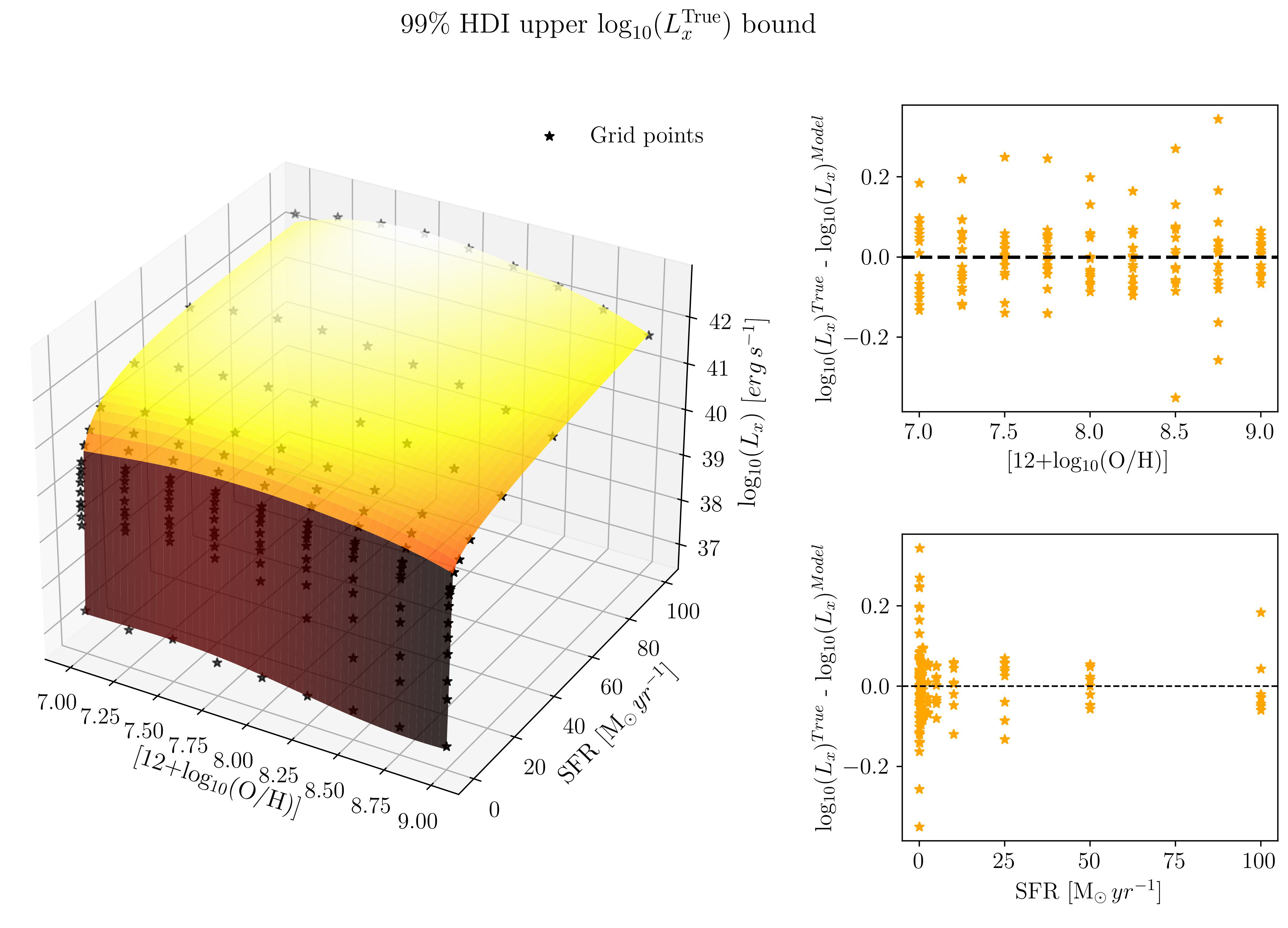} \\
\includegraphics[width=0.5\textwidth, trim={5cm 2cm 0cm 0cm},clip]{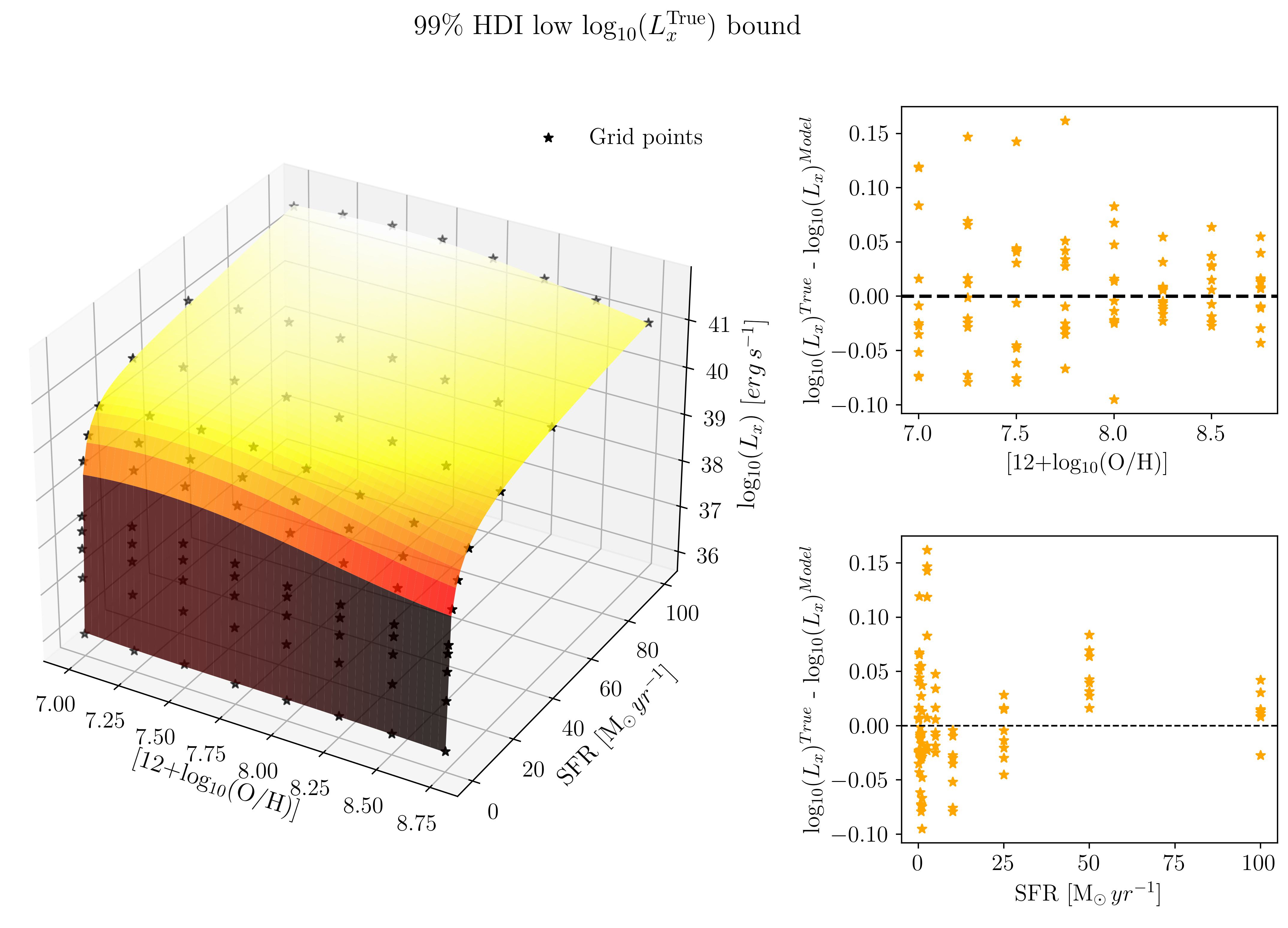}
\caption{ Distributions of the upper and lower \lx{} bounds of the $99\%$ HDI as a function of the \sfr{} and $[12+\log(\mathrm{O/H})]$ (black stars). The surfaces show the best-fit results of the upper and lower \lx{} bounds of the $99\%$ HDI together with their residuals (orange stars) as a function of the \sfr{} and the metallicity. The corresponding fit surfaces for the $68\%$, $90\%$, and $99.9\%$ HDIs are presented in Appendix~\ref{append:best-fit surfaces}.}
\label{fig:upper_lower_surface_99}
\end{figure}

\subsection{Fitting of the stochastic scatter as a function of the \sfr{} and metallicity}\label{subsec:mcmc fit}
The HDIs derived in Sect.~\ref{subsec:lx_distributions} provide a quantitative measurement of the dispersion of the X-ray luminosity distributions. Consequently, establishing correlations between the upper and lower \lx{} bounds at each interval as a function of the \sfr{} and metallicity allows us to probe the dependence of stochastic effect on the HMXB population luminosity.

In Fig.~\ref{fig:upper_lower_surface_99} we show the upper and lower \lx{} bounds at the $99\%$ HDI as a function of the \sfr{} and $[12+\log(\mathrm{O/H})]$, while the corresponding distributions for the $68\%$, $90\%$, and $99.9\%$ HDIs are given in Appendix~\ref{append:best-fit surfaces}. As indicated by the data, the lower and upper \lx{} bounds in each confidence interval, as a function of \sfr{} and metallicity, define an \lx{}–\sfr{}–($\rm [12+log(O/H)]$) surface. We therefore fitted the upper and lower limits separately with a fifth-order polynomial of two variables:
\begin{equation}
\rm y(X,Y|c_{ij}) = \sum_{i=0}^{n}\sum_{j=0}^{n-i} c_{ij} \cdot X^{i} \cdot Y^{j}, \quad n=0,\dots5,
\end{equation}
where $\rm X = [12+log(O/H)]$, $\rm Y = log_{10}{SFR}$, and $\rm c_{ij}$ is the polynomial coefficient. We note that prior to adopting the fifth-order polynomial, we tested second to sixth order polynomials. Our selection was based on two criteria: (i) the goodness of fit, determined by minimizing the Root-Mean-Squared Error (RMSE) for the upper 99\% and 99.9\% HDIs, as these extreme cases significantly impact the upper end of the HMXB XLF and the subsequent discrimination between normal galaxies and AGN, and (ii) a qualitative evaluation of the residuals in the \lx{}–\sfr{} and \lx{}–[12+log(O/H)] projections. Our analysis demonstrated that the RMSE is minimized  when using a fifth-order polynomial. Furthermore, the fit residuals confirmed that a fifth-order polynomial ensures the optimal balance between accuracy in the low-\sfr{} regime and overall model stability, avoiding the unnecessary complexity or instabilities found in higher-order models.

The complex shape of the \lx{} distributions, as well as the high dimensionality of the polynomial make standard optimisation techniques prone to degeneracies or convergence to local minima. For that reason, in our analysis, all eight surfaces were fitted with the Markov chain Monte Carlo (MCMC) technique using the Python \texttt{emcee} package \citep{emcee}. This approach offers a more robust exploration of the parameter space while providing posterior distributions for all coefficients, and hence statistically well-defined uncertainties on the fitted surfaces. The likelihood function for each surface is: 
\begin{equation}
p(y | \rm X, Y) =  -\frac{N}{2}ln(2\pi \sigma_{y_{\rm D}}) - \frac{1}{2}\sum_{n}\frac{(y_{\rm D}- y_{n}(X,Y|c_{ij}))^{2}}{2\sigma_{y_{\rm D}}^{2}}\,\,\, n=1,..N,
\end{equation}
where $\rm y_{D} = log_{10}L_{X,~upb~or~lowb}^{HDI}$, and HDI corresponds to the measured $68\%$, $90\%$, $99\%$, and $99.9\%$ intervals for each upper and lower bound. N is the number of the grid points ($N=144$). We assumed a constant uncertainty of $\sigma_{y_{\rm D}} = 0.1 \, \rm dex$ for each data point. For the MCMC fitting we used 100 walkers over 10\,000 iterations, while the first 2\,000 are discarded as part of the burn-in phrase where convergence is achieved.

Given the high dimensionality of the parameter space (21 free parameters), an adequate guess of the initial model parameters is essential to ensure convergence. For that reason, we performed a pre-fitting analysis using the \texttt{curve\_fit} package from the \texttt{scipy} Python library \citep{virtanen20}, and adopted the resulting best-fit values as initial conditions for the MCMC sampling. A uniform non-informative prior was adopted covering a wide range for each parameter. 
In an initial attempt, we allowed all 21 parameters to vary freely, but the fit did not converge. To address this, we utilised the covariance matrix from the pre-fitting analysis to calculate the relative uncertainty of each parameter. Parameters with relative errors larger than 100\% were fixed at their best-fit values, as they carried no constraining power and prevented convergence. The remaining parameters were left free during the MCMC fit. This strategy of combining a preliminary deterministic fit with selective parameter fixing is a common approach in high-dimensional polynomial models, as it stabilises the sampling and ensures robust convergence. Fig.~\ref{fig:upper_lower_surface_99} shows the best-fit surfaces of the upper and lower \lx{} bounds of the $99\%$ HDI together with their residuals as a function of the \sfr{} and the metallicity. The corresponding fits for the $68\%$, $90\%$, and $99.9\%$ HDIs are presented in Appendix~\ref{append:best-fit surfaces}. The best-fit coefficients, along with their $16\%$–$84\%$ confidence ranges, are reported in Table~\ref{tab:best-fit}. Fixed parameters are indicated by zero uncertainties. 
As shown, our best-fit models reproduce very well the $\rm log_{10}L_{X,upbor~lowb}^{HDI}$–\sfr{}–[12+log(O/H)] surfaces for all HDIs. In particular for the upper 68\% HDI the root-mean-square error (RMSE) is 0.34 dex while for the rest of the upper HDIs becomes much less ($\lesssim0.01$) dex. Similarly, for the lower bounds our best-fits reach an RMSE $\lesssim0.12$ dex for all HDIs. We also evaluated our best-models in intermediate points, different than the grid points and we found that our prescriptions perform equally well across the entire parameter space, not just at the grid nodes.

\section{Discussion}\label{sec:application}
So far, we have modelled the \lx{} scatter arising from stochastic sampling of the metallicity-dependent HMXB XLF of L21. To this end, we simulated \lx{} distributions over a wide grid of \sfr{} and metallicity values, covering the broad range of conditions observed in star-forming galaxies. By measuring statistical quantities (i.e. HDIs; see Sect.~\ref{subsec:lx_distributions}) that capture the complex shapes of these distributions, we parametrized the luminosity scatter by fitting surfaces to the upper and lower \lx{} bounds as functions of \sfr{} and gas-phase metallicity. This analysis, provides a practical set of prescriptions to calculate the expected \lx{} for a given \sfr{} and metallicity, fully accounting for stochastic effects without the need to rerun the computationally expensive sampling of the HMXB XLF for each individual galaxy. As our prescriptions are based on a fith-order polynomial of two variables, the derived \lx{} is sensitive to any extrapolation outside the grid we used for the fits. For that reason, we note here that our presrciptions are appicable only to galaxies with 0.001$\leq$\sfr{}$\leq$100 M$_{\odot}\,yr^{-1}$ and 7$\leq$[12+log(O/H)]$\leq$9. However, this limitation is not critical since at such high \sfr{}s (i.e. \sfr{}$>$100 M$_{\odot}\,yr^{-1}$) the impact of stochasticity on the integrated X-ray luminosity becomes negligible. Furthermore, these prescriptions are intended for use in the regime where HMXBs dominate the total X-ray output. In typical star-forming galaxies, the contribution from diffuse hot gas is expected to be minor \citep[$\sim 12\%$;][]{kyritsis25}, while the contribution from LMXBs remains secondary as long as the specific star-formation rate (sSFR=\sfr{}/\stellarmass{}) remains high (sSFR $ \gtrsim 10^{-10} \, \text{yr}^{-1}$). Consequently, our model is best suited for star-forming systems where the X-ray emission is driven by young stellar populations. For galaxies with significantly lower sSFRs, the LMXB population may contribute a non-negligible fraction to the total $\rm L_X$, and the stochasticity of the LMXB XLF should be also considered in the simulations, which is beyond the scope of this work. In the following sections, we demonstrate the applicability of our prescriptions by exploring the implications of stochasticity for both local and high-redshift galaxy populations.

\subsection{Implications of the stochastic sampling of the HMXBs XLF on the individual galaxies}\label{subsed:individuals}

\begin{figure}
        \includegraphics[width=\columnwidth]{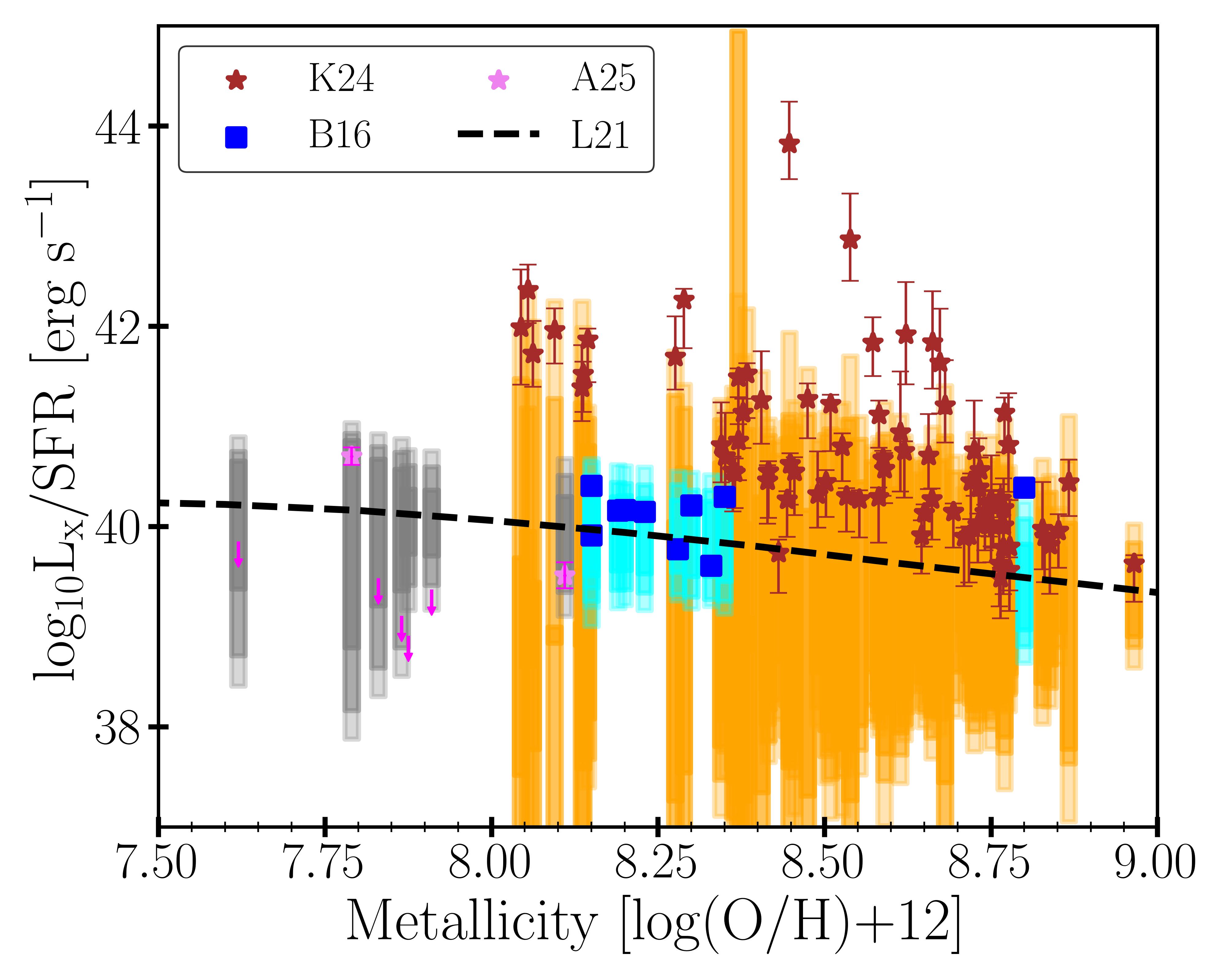}
    \caption{The \lx{}/\sfr{} ratio as a function of metallicity. The upper and lower $90\%$, $99\%$, and $99.9\%$ HDI bounds of the expected \lx{} from stochastic sampling are shown as shaded bands of different colours for each data set. Violet stars with grey-shaded regions correspond to the sample of \citet{adamcova24}, brown stars with orange-shaded regions to \citet{kyritsis25}, and blue squares with cyan-shaded regions to \citet{brorby16}. For reference, the \lx{}/\sfr{}–[12+log(O/H)] scaling relation of L21 is shown as a black dashed line.}
    \label{fig:lx_SFR_metal}
\end{figure}

As discussed in Sect.~\ref{sec:intro}, stochastic sampling of the HMXB XLF can lead to elevated integrated X-ray emission in normal galaxies, as well as to increased scatter around the standard \lx{}-\sfr{}-Metallicity scaling relations. Consequently, exploring the role of stochasticity is a necessary step before attributing deviations in the scaling relations to rare source populations, variations in binary formation channels and efficiency, or to contribution from AGN activity. Our prescriptions provide a practical framework for quantifying the expected \lx{} due solely to stochastic sampling of the HMXB XLF, without the need to rerun the full suite of MC simulations for each galaxy. To demonstrate their applicability, we applied our prescriptions to the datasets analysed in two recent studies, \citet{adamcova24} and \citet{kyritsis25}, which examined whether stochastic sampling of the HMXB XLF could account for the increased scatter observed in their sample of star-forming galaxies. In Fig.~\ref{fig:lx_SFR_metal} we present the results of our analysis in terms of the \lx{}/\sfr{} ratio as a function of metallicity. The upper and lower 90\%, 99\% and 99.9\% HDI bounds of the expected \lx{} due to stochastic sampling are shown as shaded bands with different colours. For reference we also plot the \lx{}/\sfr{} - [12+log(O/H)] scaling relation from L21. Our results confirm the conclusions of these works that stochasticity alone cannot explain the full amplitude of the observed scatter. 

In particular, in their work \cite{adamcova24} studied a set of blueberry galaxies, which are considered local analogues of the compact star-forming galaxies dominating the early Universe \citep[e.g.;][]{kouroumpatzakis24}. They found large scatter in X-ray properties, with one source exhibiting enhanced X-ray luminosity while others appeared underluminous. As shown in their analysis, as well as, in Fig.~\ref{fig:lx_SFR_metal} this large scatter cannot be fully explained by stochasticity effects. Specifically, the brightest source in their blueberry galaxy sample (BB8), lies above the expected distribution at nearly the 90\% HDI upper bound, suggesting the possible presence of an additional X-ray emitter such as a hidden LLAGN or an XRB with extreme X-ray luminosity. Conversely, the second detected source (BB1) lies more than 90\% below the lower bounds of the stochastic sampling, while two further sources, detected only as X-ray upper limits (BB3 and BB4), fall more than 99.9\% below the predicted stochastic lower bounds. As discussed in their work, this X-ray deficit could reflect the extremely young starburst nature of these galaxies, in which stellar populations are not yet old enough to form a significant population of XRBs.

Similarly, in the first all-sky survey of star-forming galaxies with eROSITA \citep{predehl21}, \cite{kyritsis25} found a significant population of galaxies with elevated integrated X-ray emission and much larger scatter around the standard scaling relation observed in individual galaxies. As illustrated, in Fig.~\ref{fig:lx_SFR_metal}, this population lies above the 99.9\% HDI upper bound expected from stochastic sampling, and in some cases almost 2 dex above the standard L21 scaling relation. As discussed, by \cite{kyritsis25}, this excess is strongly correlated with lower metallicity galaxies, but the primary driver is the age of the stellar populations. This led to the identificantion of a sub-population of very X-ray luminous starburst galaxies characterized by high specific star formation rates, lower metallicities, and younger stellar populations, which effectively drives upwards the X-ray luminosity, increasing the observed scatter.

The above analysis demonstrates how our prescriptions can be applied to disentangle the role of stochasticity from intrinsic astrophysical effects in very large samples, without the need for customized simulations for each individual object. In this way, they provide a necessary first step in assessing whether observed X-ray excesses or deficits in star-forming galaxies are due to stochastic sampling or instead reflect fundamental differences in the astrophysical nature of the X-ray emission origin.

\subsection{Redshift evolution of the stochastic scatter of the X-ray luminosity }\label{subsec:z-evolution}

Previous studies have found that the main driver of the elevated \lx{}/\sfr{} ratios observed in local analogues of high-redshift galaxies, as well as in high-redshift galaxies, is their lower metallicity. Low-metallicity environments result in the formation of more massive compact objects and XRBs with shorter orbital separations \citep{linden10,fragos13a,brorby16,basuzych16,fornasini18}, which in turn produce more luminous XRBs and hence higher integrated X-ray emission. However, at higher redshifts, where galaxies are typically younger, less massive, and metal-deficient, stochastic sampling effects on an individual galaxy basis may also play an important role, contributing to the observed scatter around the \lx{}-\sfr{}-Metallicity scaling relation. In addition, such galaxy-by-galaxy stochastic fluctuations, may impact the interpretation of deep field surveys. In particular, if the sensitivity limit of a survey does not reach the average luminosity of the galaxy population, the detected sources will consist preferentially by galaxies  exhibiting positive luminosity fluctuations with respect to the mean scaling relations. This leads to a systematic bias since we probe the high-luminosity end of the HMXBs XLF while missing the bulk of the fainter populations, leading to biased scaling relations possibly overestimating the X-ray luminosity evolution. 

Investigating whether stochastic effects themselves evolve with redshift is crucial for determining whether the observed \lx{}/\sfr{} evolution can be fully explained by metallicity and stellar-population age effects \citep{gilbertson22}, or whether the observed populations are at odds with the current scaling relations pointing to the emergence of superluminous XRB populations or additional formation channels that are not accounted for in the current XLF models. 

\begin{figure}[t]
        \includegraphics[width=\columnwidth]{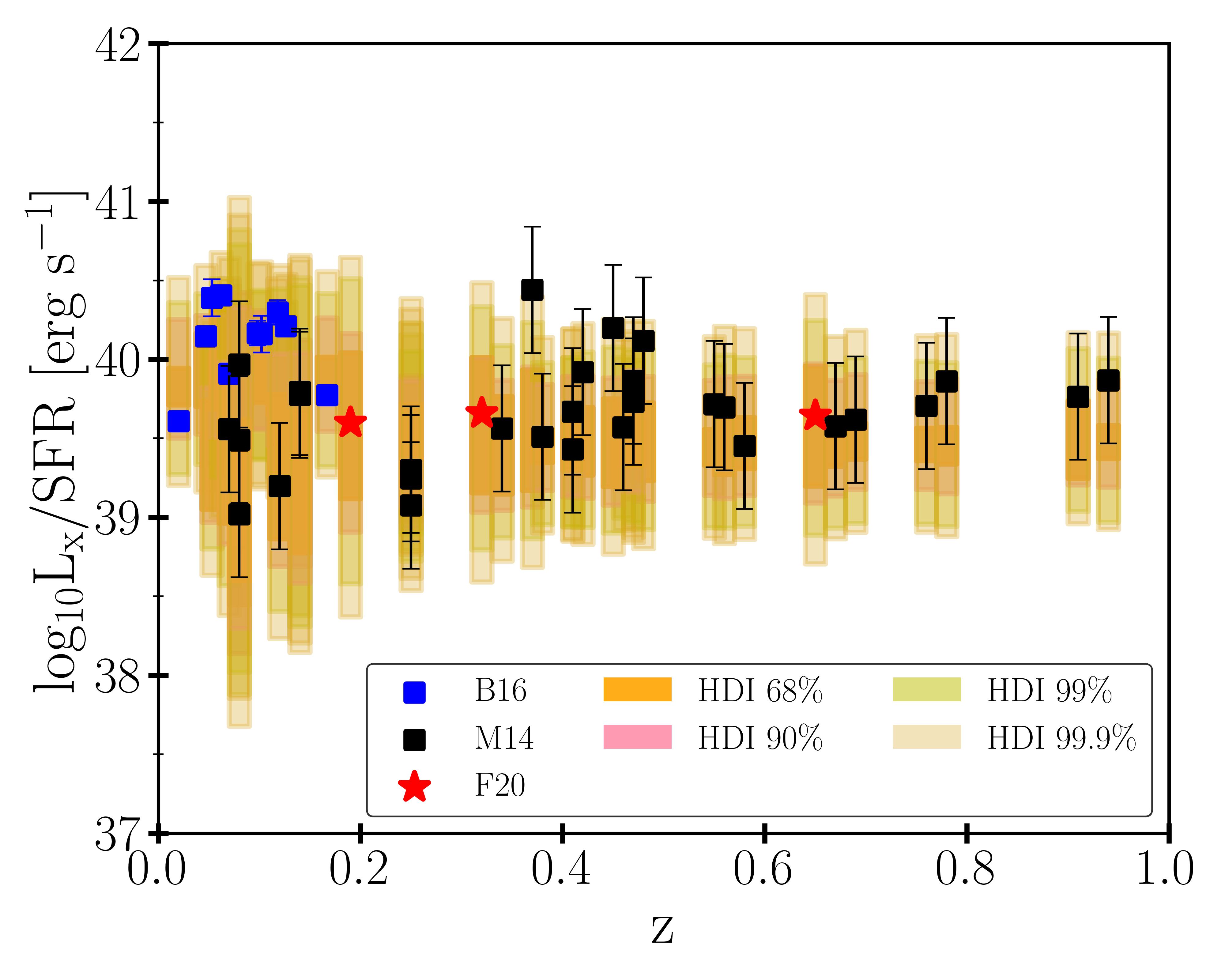}
    \caption{The \lx{}/\sfr{} ratio as a function of redshift. The upper and lower $68\%$, $90\%$, $99\%$, and $99.9\%$ HDI bounds of the expected \lx{} from stochastic sampling are shown as shaded bands of different colours. The majority of the objects are consistent with the scaling relations which accounts for the effect of metallicity.  Blue squares correspond to the sample of \citet{brorby16}, while black stars to the sample of \cite{mineo14}. Red stars show the stacked data from \cite{fornasini20}}
    \label{fig:lx_SFR_redshift_evolution_indiv}
\end{figure}

As an example, we apply our prescriptions to two samples of star-forming galaxies spanning the redshift range z$\simeq$0-1. In Fig.~\ref{fig:lx_SFR_redshift_evolution_indiv} we show their \lx{}/\sfr{} ratios as a function of redshift, together with the corresponding upper and lower 68\%, 90\%, 99\%, and 99.9\% HDI bounds of the expected \lx{} due to stochastic sampling, which are represented as shaded bands of different colours. The first sample (blue squares) is from the work of \citeauthor{brorby16} \citeyear{brorby16} (B16), who studied the X-ray emission of HMXBs for a sample of Lyman break analogue (LBA) galaxies, which serve as local examples of early, metal-poor galaxies. The second sample (black squares) is from the work of \citeauthor{mineo14} \citeyear{mineo14} (M14) who investigated the relation between the integrated X-ray emission from star-forming galaxies and their \sfr{} by using a sample of high-redshift galaxies from Chandra Deep Fields (CDFs). Since \cite{mineo14} do not report metallicities, we estimated the gas-phase metallicities for their sample using the fundamental mass-metallicity relation from \cite{curti20}, and the stellar masses provided in M14.

As shown in Fig.~\ref{fig:lx_SFR_redshift_evolution_indiv}, at lower redshifts (z$<0.25$) the scatter (indicated by the length of the bands) is larger, while at higher redshifts it becomes smaller. This trend arises because the higher-redshift galaxies detected in X-ray surveys typically exhibit systematically higher \sfr{}s, which leads to the formation of larger HMXB populations, thereby reducing the role of the stochastic effects. In contrast, local galaxies generally have lower \sfr{}s, where stochasticity dominates and increases the observed \lx{}/\sfr{} scatter. In addition, the \lx{}/\sfr{} ratios of the B16 galaxy sample are systematically elevated compared to those of M14. This offset can be attributed to the lower metallicities of the B16 galaxies and does not require the presence of new population of superluminous XRB populations or alternative formation channels. In fact the measurements are within the 90\% HDI bounds estimated after taking into account the \sfr{} and the metallicity of each galaxy. This interpretation is further supported by Fig.~\ref{fig:lx_SFR_metal}, where, once metallicity is taken into account, the B16 galaxies closely follow the \lx{}/\sfr{}–[12+log(O/H)] scaling relation. 

Interestingly, we find that the M14 subsample drawn from the CDF surveys, exhibit X-ray luminosities generally above the 68\% HDI upper bound of the expected \lx{}, and in some cases even above the 90\% HDI. This effect becomes more pronounced at higher redshifts, suggesting that deep field surveys, constrained by their sensitivity limits, preferentially detect only galaxies whose X-ray emission is dominated by the bright end of the HMXBs XLF. As a consequence, these surveys sample the most luminous star-forming galaxies while missing the bulk of the galaxy populations with lower X-ray luminosities. This interpretation is supported by the stacked X-ray measurements of \cite{fornasini20} (red stars), which probe the average population of star-forming galaxies (including undetected X-ray galaxies) in three redshift bins. Their results lie much closer to the mean \lx{} values predicted by the HMXB XLF, indicating that stacking recovers the typical galaxy population rather than positive fluctuations of a few bright sources.

The importance of quantifying these stochastic effects is highlighted also by recent works which have demonstrated that ignoring the galaxy-to-galaxy stochastic scatter around the \lx{}/\sfr{}–Metallicity relation, which originates from the stochastic nature of binary production and their X-ray luminosity, may lead to biased estimates of the cosmic X-ray emissivity, especially in high redshifts \citep{nikolic24}. 

To obtain a more complete view of whether stochastic scatter evolves with redshift, we carried out a simulation study. In this framework, we drew the galaxy properties (i.e. \sfr{}, [12+log(O/H)]) at different redshifts and by employing our prescriptions we calculated the expected \lx{} due to stochastic sampling of the HMXB XLF across different redshift bins. In this simulation study we aim to measure the expected scatter, and compare it with the redshift-dependent \lx{}–\sfr{} scaling relation of 
\citeauthor{lehmer16} \citeyear{lehmer16} (L16; third equation in their Table 3; 2-10 keV energy band) as an indicative reference. 
While the choice of the band does not affect the result of the stochasticity, it does affect the absolute offset from the measured luminosities. For this reason we converted the L16 scaling relation from the $2-10$ keV band into the $0.5-8$keV band adopted in our work using the  spectral models of starburst galaxies presented in \cite{lehmer15}. The comparison will enable us to assess whether there is intrinsic excess that should be taken into account in the analysis of the scaling relations. 

As a first step, we defined a set of redshifts, z$ = [0.5, 0.8, 1.1, 1.5, 2, 2.5, 3, 3.5, 4, 5]$, and for each of these we simulated 100 galaxies with different \stellarmass{} drawn from the stellar mass functions (SMFs) of \citet{davidzon17}, who provided consistent estimates of the SMF in the COSMOS field between z$=0.1$ and 6, based on near-IR data from the COSMOS2015 catalogue \citep{laige16}. For each redshift bin, we randomly sampled 100 galaxy stellar masses in the $10^{7} \leq$ \stellarmass{} $\leq 10^{12}$ M$_\odot$ mass range, from the corresponding best-fitting Schechter function \citep[active galaxy sample from Table 1 of][]{davidzon17}. The corresponding \sfr{} values were then derived using the redshift-dependent star-forming galaxy main sequence (SFMS) from \cite{speagle14}, who compiled and homogenized results from 25 different SFMS studies from the literature and fitted its redshift evolution. Finally, gas-phase metallicities were assigned following the redshift-dependent mass-metallicity relation of \cite{nikolic24} which is based on the fundamental mass-metallicity relation from \cite{curti20}. In the derivation of the \sfr{}s and metallicities, we accounted for the intrinsic scatter of both the SFMS and the fundamental mass-metallicity relation, by adding Gaussian noise, with a standard deviation corresponding to the intrinsic uncertainties, reported by \cite{speagle14} ($\sim 0.3$~dex) and \cite{curti20} ($\sim 0.054$~dex), respectively.
Using the \sfr{} and [12+log(O/H)] values of the 100 simulated galaxies in each redshift bin, we applied our prescriptions to calculate the expected \lx{} within different HDIs. Then we calculated the excess of these HDIs relative to the redshift-dependent \lx{}–\sfr{} scaling relation of L16, defined as:
\begin{equation}
    \rm Excess\,from \, L16 = log_{10}L_{X,~upb~or~lowb}^{HDI} - log_{10}L_{X}^{L16},
\end{equation}
where the L$\rm _{X,~upb}^{HDI}$ and L$\rm _{X,~lowb}^{HDI}$ indicate the upper and lower bound of the different HDI as defined in Sect.~\ref{subsec:lx_distributions}.

\begin{figure*}
        \includegraphics[width=0.95\textwidth]{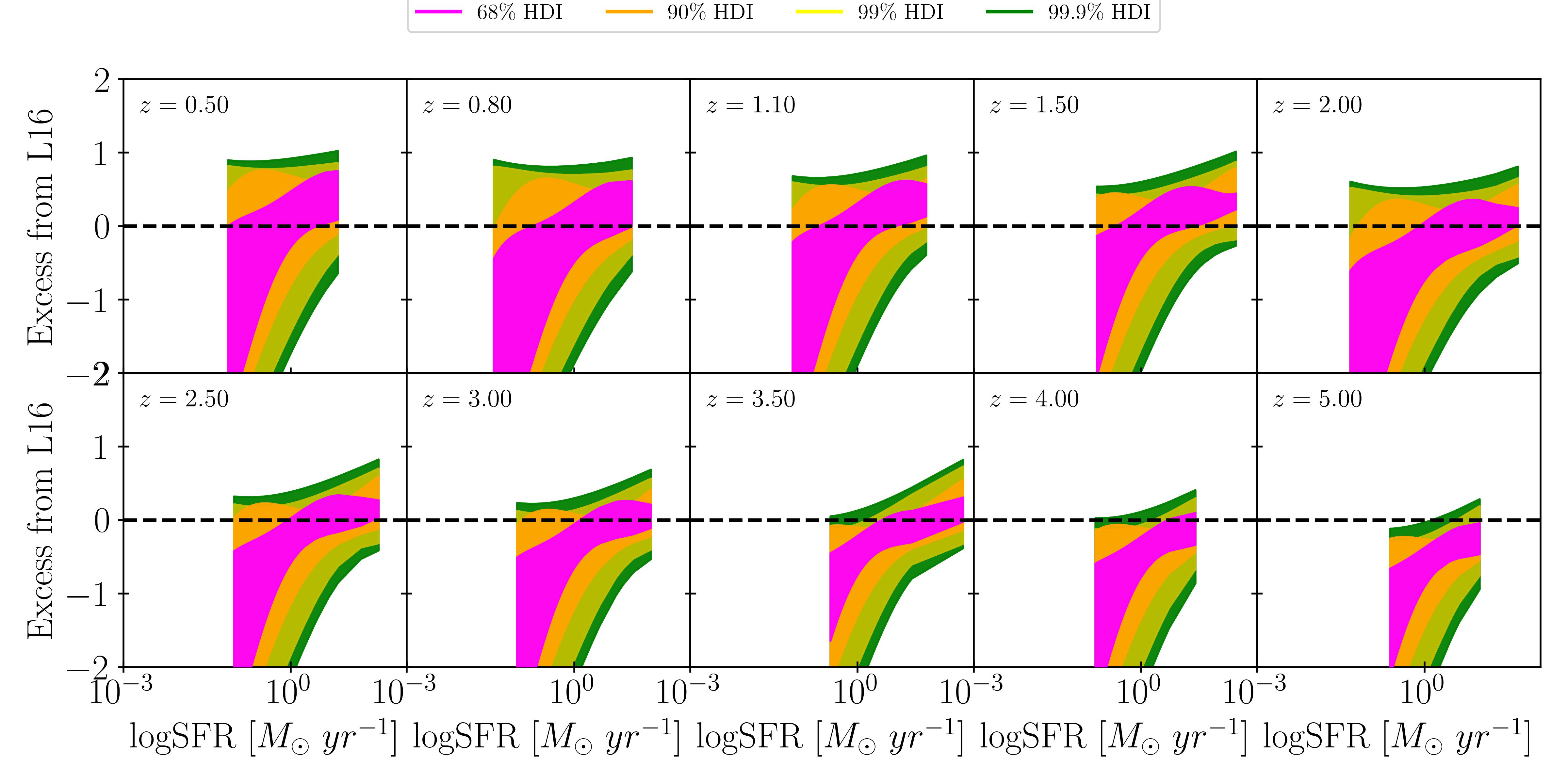}
    \caption{The \lx{} excess, defined as log$_{10}$L$_{X,~upb~or~lowb}^{HDI}$ - log$_{10}$L$_{X}^{L16}$ as a function of \sfr{} for all the redshift bins considered in this simulation study. The various HDIs indicated by shaded areas of different colours. The black dashed line indicates the line of equality between the \lx{} from stochastic sampling and the \lx{} from the redshift-dependent scaling relation of \cite{lehmer16} (L16). The shift of the maximum \sfr{} in each redshift bin, reflects the \sfr{} distribution of the galaxies in that bin.}
    \label{fig:lx_excess_redshift_evolution}
\end{figure*}

In Fig.~\ref{fig:lx_excess_redshift_evolution}, we plot this \lx{} excess as a function of \sfr{}, with the various HDIs indicated by shaded areas of different colours. The black dashed line indicates the line of equality between the \lx{} from stochastic sampling and the \lx{} from L16. The L16 relation traces the integrated \lx{} of star-forming galaxies as a function of \sfr{} across different redshift bins based on observations from local galaxies and deep surveys. Since it uses low-redshift galaxies (z$<0.5$), it asymptotically converges to the average \lx{} for local galaxies which were used for the derivation of the L21 scaling relation employed in our prescriptions. The magnitude and the asymmetry of the excess with respect to the line of equality quantifies the strength of the scatter around this mean. This allows us to assess the impact of stochastic scatter as a function of redshift.

The first thing that becomes clear from Fig.\ref{fig:lx_excess_redshift_evolution}, is that in the first five redshift bins (z$\lesssim2$) the excess is asymmetric with respect to the middle of the 68$\%$ HDI and highly skewed to higher luminosities regardless of the \sfr{}. This results in \lx{} nearly 1 dex above the L16 scaling relation, which is far higher than the $\sim0.25$ dex intrinsic scatter of that relation. Focusing on the 99\% and 99.9\% upper envelopes of the excess, we find that they remain approximately constant across the full range of \sfr{} values up to z$\lesssim0.8$, while in higher redshift (z$\sim1-2$) there is a systematic increase of the upper envelope with respect to the mean. The assymetric behavior with respect to the mean as a funtion of \sfr{} can be interpreted if we consider Fig.\,~\ref{fig:samples_distribution} where we see that in low \sfr{}s there is a bump at high \lx{} which disappears at higher \sfr{}s while the mean of the \lx{} distribution remains unchanged. Therefore, the lower \sfr{} galaxies in the lower redshifts will result in stronger stochasticity and increased integrated luminosities with respect to the mean (black line in Fig.\,~\ref{fig:samples_distribution}). As we see from Fig.\,~\ref{fig:lx_excess_redshift_evolution} the difference between the expected integrated \lx{}, derived from the scaling relation of L16, and the upper 99.9\% HDI remains almost constant from \sfr{}$\sim0.1 \, \rm M_{\odot}\,yr^{-1}$ to $1 \rm M_{\odot}\,yr^{-1}$ for solar metallicities resulting in a constant upper envelope of the scatter at the lower redshift bins. At higher redshifts the galaxies have higher \sfr{}s for a given \stellarmass{} (based on the redshift evolution of the SFMS). In addition, the SMF becomes steeper in higher redshifts, but without a significant change in its upper bound. This way the wide range of stellar mass results in a wide range of metallicity, which at the highest redshifts is skewed towards lower metallicity as a result of the redshift evolution of the mass-metallicity relation \citep{curti24}. 
As a result, at higher z-bins galaxies of a given \sfr{} are associated with galaxies with lower \stellarmass{} (due to the redshift-evolution of the SFMS), and hence lower metallicity. From Fig.~\ref{fig:samples_distribution} we see that the systematically lower metallicities lead to a higher upper envelope with respect to the mean. This combined with the higher \sfr{} and lower metallicities (on average) in these bins, results in strong evolution of the upper envelope in Fig.~\ref{fig:lx_excess_redshift_evolution} as a function of \sfr{}. 

At redshifts z=2.5 and z=3, which correspond to the peak of the cosmic star-formation history \citep{madau14}, the behaviour of the excess is similar with a consistently increasing upper envelope in high \sfr{}, and a wider low envelope in low \sfr{}. However, we now see that in higher \sfr{} the excess is more symmetric around the L16 relation than in the lower z bins. In these redshift bins the \sfr{}s are sufficiently high to produce nearly Gaussian distributions of X-ray luminosity with relatively small scatter around the mean (Fig.~\ref{fig:samples_distribution}). At very high redshifts (z=4–5), the expected \lx{} from stochastic HMXB sampling falls below the L16 prediction. This is possibly due to the fact that the measured \lx{} scalling relations at these redshifts are based on scarce samples probing the most extreme galaxies, indicating the bias of flux-limited surveys towards more actively star-forming and more X-ray luminous galaxies at higher redshifts.

\begin{figure}
        \includegraphics[width=\columnwidth]{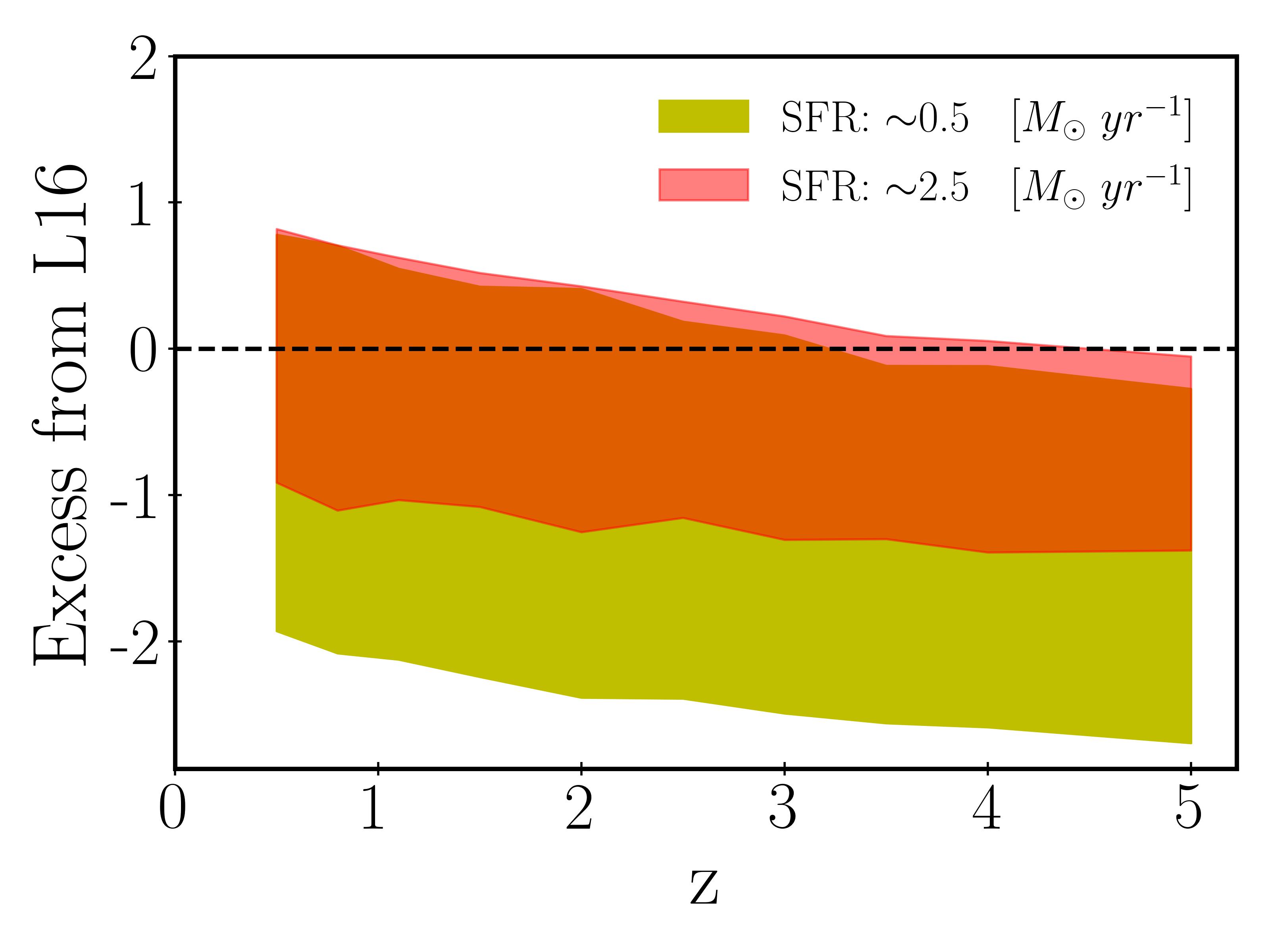}
    \caption{The \lx{} excess at the 99\% \lx{} bounds from the L16 as a function of redshift for two indicative \sfr{} values: 0.5 $\rm M_{\odot}/yr$ (yellow shaded area) and 2.5 $\rm M_{\odot}/yr$ (orange shaded area)}
    \label{fig:lx_SFR_stoch_scatter}
\end{figure}

Overall, our results indicate a weak decrease of the intrinsic scatter in a given \sfr{}, up to z$=2.5$. To better trace the redshift evolution of the stochastic \lx{} scatter at fixed \sfr{}, Fig.~\ref{fig:lx_SFR_stoch_scatter} shows the \lx{} excess at the 99\% upper and lower bounds relative to the L16 relation as a function of redshift for two indicative \sfr{} values: 0.5~$\rm M_{\odot},yr^{-1}$ (low \sfr{}; yellow shaded area) and 2.5~$\rm M_{\odot},yr^{-1}$ (moderate \sfr{}; orange shaded area). We find that for the lower \sfr{} the scatter is systematically larger across all redshifts compared to the higher-\sfr{} case. This again reflects the stronger stochastic sampling effects in low–star-formation environments, where the flat HMXB XLF leads to more unstable sampling. More importantly, at high redshifts ($\rm z>4$) the 99\% upper envelope of the \lx{}/\sfr{} scatter falls below the expected \lx{} predicted by the L16 scaling relation. This is probably due to the fact that in the highest redshifts the L16 relation is based on the higher \sfr{} and higher \lx{} galaxies, thereby missing the bulk of the  intrinsically X-ray fainter galaxy populations at these redshifts. 

As discussed in Sect.~\ref{sec:intro}, another important application of quantifying the stochastic scatter in the X-ray luminosity output of HMXBs in galaxies is assesing the presence of AGN based on the X-ray emission alone or when other signatures (e.g. optical or infrared) are absent. Our results suggest that at z$\lesssim1.1$ the \lx{} of stochastically sampled HMXB populations can exceed the L16 prediction by up to 1 dex. In this regime, there is significant overlap between the X-ray luminosities of star-forming galaxies and those typically attributed to LLAGN. At higher redshifts (1.5$\lesssim$z$\lesssim$3) the effect is less pronounced, though contamination from highly star-forming, low-metallicity galaxies remains possible. This implies that samples of AGN in deep surveys selected solely on the basis of their X-ray luminosity may be subject to contamination by luminous star-forming galaxies in the low-end of the AGN X-ray luminosity function. The presented prescriptions can be used to quantify what is the probability for a given galaxy to have as high luminosity as that measured for a candidate AGN in a deep survey.

\section{Conclusions}\label{sec:conclusion}
In this work we investigated the role of stochastic sampling of the HMXB XLF in the integrated \lx{} of star-forming galaxies. To this end, we simulated \lx{} distributions over a wide grid of \sfr{} and metallicity values, covering the range of conditions observed in local and higher-$z$ galaxies. By measuring statistical quantities (e.g. HDIs; see Sect.~\ref{subsec:lx_distributions}) we parametrised the luminosity scatter by fitting surfaces to the upper and lower \lx{} bounds as a function of \sfr{} and gas-phase metallicity. Our analysis provides a practical set of prescriptions for calculating the expected \lx{} for a given \sfr{} and metallicity within the \sfr{}: 0.001-100 M$_{\odot}\,yr^{-1}$ and [12+log(O/H)]:7-9 range, fully accounting for stochastic effects without the need to rerun the computationally expensive sampling of the HMXB XLF for each individual galaxy.

By applying these prescriptions to both local and higher-redshift galaxy samples, we showed that stochasticity must be considered before attributing the observed X-ray emission of star-forming galaxies to intrinsic differences in their underlying populations. Furthermore, a galaxy simulation study across $z=0.5$–5 showed mild intrinsic redshift evolution of the stochastic scatter. Its behaviour mirrors the evolution of \sfr{} and metallicity, with minimum scatter at $z\simeq 2.5$ where the cosmic star-formation density reached its maximum. In addition, our analysis of the \lx{}/\sfr{} excess shows that the intrinsic scatter is larger in lower-SFR galaxies, while it decreases in higher-SFR systems, where the shape of the HMXB XLF is primarily driven by the metallicity. Interestingly, at high redshifts ($z>4$), the 99\% upper envelope of the \lx{}/\sfr{} excess lies below the \lx{} predicted by the L16 scaling relation. This behaviour suggests that current redshift-dependent scaling relations are likely biased toward more luminous galaxies that occupy the bright end of the HMXB XLF. Consequently, these relations, which are derived from flux-limited deep X-ray surveys, may fail to capture the bulk of the HMXB population at these epochs, which is intrinsically fainter. Furthermore, we find that the upper 99\%–99.9\% bounds of the stochastic \lx{} overlap with the luminosity regime typically associated with LLAGN (\lx{} $\gtrsim 10^{41}$ erg s$^{-1}$). This overlap is most pronounced at $z<1.1$, but remains non-negligible at intermediate redshifts. Consequently, deep-field surveys that neglect stochastic effects risk misclassifying luminous star-forming galaxies as AGN, thereby biasing AGN demographics.

Overall, our results demonstrate that stochastic sampling of the HMXB XLF is a fundamental source of scatter in the \lx{}/\sfr{} relation, especially at low \sfr{}s. Accounting for these effects is essential for quantifying potential sources of bias in cosmological studies of the X-ray emission of galaxies, and disentangling normal star formation from AGN activity in deep surveys. The prescriptions presented here offer a practical framework for constraining uncertainties in scaling relations, quantifying the likelihood of extreme outliers, refining the classification of X-ray sources in current and future surveys, and estimating the expected X-ray luminosity from high-z galaxies.

\section{Data availability}\label{sec:data_availability}
The code for the application of our prescriptions, the best-model \lx{}-\sfr{} projection plots (for fixed metallicity bins), the best-model \lx{}-Metallicity projection plots (for fixed \sfr{} bins), and the \lx{} draws used in this analysis, are publicly available via GitHub repository in the following link: \url{https://github.com/EliasKyritsis/Lx\_stochasticity\_prescriptions}.

\begin{acknowledgements}
We thank the anonymous referee for comments that helped to improve the clarity of the manuscript. This work was supported by the \emph{Found\-ation for Re\-search \& Tech\-nology – He\-llas} (FORTH). EK  acknowledges support from the \emph{FORTH Synergy Grant PARSEC}, and from the Public Investments Program through a Matching Funds grant to the IA-FORTH. The research leading to these results has received funding from the European Union’s Horizon 2020 research and innovation programme under the Marie Skłodowska-Curie RISE action, Grant Agreement n. 873089 (ASTROSTAT-II). KK was supported an ICE Fellowship under the program Unidad de Excelencia Maria de Maeztu (CEX2020-001058-M). We wish to thank the AstroStat Academy for providing training on the analysis methods adopted in this work.
\end{acknowledgements}

% WARNING
%-------------------------------------------------------------------
% Please note that we have included the references to the file aa.dem in
% order to compile it, but we ask you to:
%
% - use BibTeX with the regular commands:
%   \bibliographystyle{aa} % style aa.bst
%   \bibliography{Yourfile} % your references Yourfile.bib
%
% - join the .bib files when you upload your source files
%-------------------------------------------------------------------
\bibliographystyle{aa}
\bibliography{references}

\begin{appendix}

\section{Best-fit surfaces for the 68\%, 90\%, and 99.9\% HDIs}\label{append:best-fit surfaces}
In this Appendix we present the best-fit surfaces of the upper and lower \lx{} bounds of the $68\%$, 90\%, and 99.9\% HDIs together with their residuals as a function of the \sfr{} and the metallicity. The best-fit coefficients, along with their $16\%$–$84\%$ confidence ranges, are reported in Table~\ref{tab:best-fit}. Fixed parameters are indicated by zero uncertainties. 

\begin{table*}
\centering
\caption{Best-fit parameters of the $\rm log_{10}L_{X,~upb~or~lowb}^{HDI} = y(X,Y)^{\ast}$.}
\label{tab:best-fit}
\begin{tabular}{ccccc}
\hline\hline
Parameter & HDI 68\% & HDI 90\% & HDI 99\% & HDI 99.9\% \\
\hline
\multicolumn{5}{c}{Upper-Bound} \\
\hline
c00 & -1080.16$^{+0.0}_{-0.0}$ & -1111.38$^{+0.0}_{-0.0}$ & 3193.43$^{+0.0}_{-0.0}$ & 1644.59$^{+0.0}_{-0.0}$ \\
c01 & 94.63$^{+0.0}_{-0.0}$ & 379.88$^{+0.0}_{-0.0}$ & -229.48$^{+0.0}_{-0.0}$ & -93.65$^{+0.0}_{-0.0}$ \\
c02 & -2.88$^{+0.0}_{-0.0}$ & -11.53$^{+0.0}_{-0.0}$ & -22.37$^{+0.0}_{-0.0}$ & 1.99$^{+0.0}_{-0.0}$ \\
c03 & 0.89$^{+0.0}_{-0.0}$ & -3.2027$^{+0.3186}_{-0.3217}$ & -3.0328$^{+0.1889}_{-0.1881}$ & 0.8195$^{+0.1891}_{-0.1912}$ \\
c04 & -0.1605$^{+0.0187}_{-0.0191}$ & -0.3338$^{+0.0146}_{-0.0147}$ & -0.2898$^{+0.0194}_{-0.0193}$ & 0.0008$^{+0.0248}_{-0.0249}$ \\
c05 & 0.0758$^{+0.0039}_{-0.0039}$ & -0.0382$^{+0.0036}_{-0.0038}$ & 0.0032$^{+0.0019}_{-0.0018}$ & 0.0012$^{+0.0019}_{-0.0019}$ \\
c10 & 727.97$^{+0.0}_{-0.0}$ & 717.15$^{+0.0}_{-0.0}$ & -1991.75$^{+0.0}_{-0.0}$ & -1029.67$^{+0.0}_{-0.0}$ \\
c11 & -40.71$^{+0.0}_{-0.0}$ & -188.5$^{+0.0}_{-0.0}$ & 116.86$^{+0.0}_{-0.0}$ & 43.66$^{+0.0}_{-0.0}$ \\
c12 & 1.87$^{+0.0}_{-0.0}$ & 5.02$^{+0.0}_{-0.0}$ & 8.58$^{+0.0}_{-0.0}$ & -1.0$^{+0.0}_{-0.0}$ \\
c13 & -0.2$^{+0.0}_{-0.0}$ & 0.9303$^{+0.0806}_{-0.0799}$ & 0.7173$^{+0.0463}_{-0.047}$ & -0.24$^{+0.0488}_{-0.0479}$ \\
c14 & 0.014$^{+0.0028}_{-0.0027}$ & 0.0367$^{+0.0022}_{-0.0022}$ & 0.0344$^{+0.0024}_{-0.0023}$ & -0.0007$^{+0.0031}_{-0.0031}$ \\
c20 & -191.88$^{+0.0}_{-0.0}$ & -180.14$^{+0.0}_{-0.0}$ & 501.53$^{+0.0}_{-0.0}$ & 262.62$^{+0.0}_{-0.0}$ \\
c21 & 6.23$^{+0.0}_{-0.0}$ & 35.12$^{+0.0}_{-0.0}$ & -22.01$^{+0.0}_{-0.0}$ & -7.54$^{+0.0}_{-0.0}$ \\
c22 & -0.34$^{+0.0}_{-0.0}$ & -0.69$^{+0.0}_{-0.0}$ & -1.0599$^{+0.0017}_{-0.0017}$ & 0.1649$^{+0.0018}_{-0.0018}$ \\
c23 & 0.0064$^{+0.0003}_{-0.0003}$ & -0.0608$^{+0.005}_{-0.005}$ & -0.0418$^{+0.0029}_{-0.0029}$ & 0.0171$^{+0.0031}_{-0.0032}$ \\
c30 & 25.55$^{+0.0}_{-0.0}$ & 22.77$^{+0.0}_{-0.0}$ & -62.97$^{+0.0}_{-0.0}$ & -33.3$^{+0.0}_{-0.0}$ \\
c31 & -0.38$^{+0.0}_{-0.0}$ & -2.91$^{+0.0}_{-0.0}$ & 1.82$^{+0.0}_{-0.0}$ & 0.5799$^{+0.0005}_{-0.0004}$ \\
c32 & 0.0184$^{+0.0}_{-0.0}$ & 0.0306$^{+0.0}_{-0.0}$ & 0.0423$^{+0.0002}_{-0.0002}$ & -0.0087$^{+0.0002}_{-0.0002}$ \\
c40 & -1.71$^{+0.0}_{-0.0}$ & -1.44$^{+0.0}_{-0.0}$ & 3.95$^{+0.0}_{-0.0}$ & 2.1$^{+0.0}_{-0.0}$ \\
c41 & 0.01$^{+0.0}_{-0.0}$ & 0.0906$^{+0.0}_{-0.0}$ & -0.0562$^{+0.0}_{-0.0}$ & -0.0169$^{+0.0001}_{-0.0001}$ \\
c50 & 0.05$^{+0.0}_{-0.0}$ & 0.04$^{+0.0}_{-0.0}$ & -0.0988$^{+0.0}_{-0.0}$ & -0.0529$^{+0.0}_{-0.0}$ \\
\hline
RMSE (dex)$^{\ast\ast}$ & 0.34& 0.17 & 0.09& 0.04\\
\hline
\multicolumn{5}{c}{Low-Bound}\\
\hline
c00 & 1931.31$^{+0.0}_{-0.0}$ & 5014.28$^{+0.0}_{-0.0}$ & 96.05$^{+0.0}_{-0.0}$ & 8860.18$^{+0.0}_{-0.0}$ \\
c01 & 507.12$^{+0.0}_{-0.0}$ & 194.21$^{+0.0}_{-0.0}$ & 15.05$^{+0.0}_{-0.0}$ & -90.76$^{+0.0}_{-0.0}$ \\
c02 & 83.91$^{+0.0}_{-0.0}$ & 76.96$^{+0.0}_{-0.0}$ & 67.43$^{+0.0}_{-0.0}$ & 120.39$^{+0.0}_{-0.0}$ \\
c03 & 3.7$^{+0.0}_{-0.0}$ & 2.9476$^{+0.6238}_{-0.6327}$ & 2.74$^{+0.0}_{-0.0}$ & -0.0$^{+0.0}_{-0.0}$ \\
c04 & 0.9082$^{+0.0798}_{-0.0799}$ & 0.5838$^{+0.1113}_{-0.1082}$ & 0.5739$^{+0.154}_{-0.1543}$ & 1.1904$^{+0.3009}_{-0.2999}$ \\
c05 & -0.0899$^{+0.0093}_{-0.0088}$ & -0.0164$^{+0.0151}_{-0.0158}$ & 0.097$^{+0.0297}_{-0.0292}$ & 0.0919$^{+0.0597}_{-0.0609}$ \\
c10 & -1253.83$^{+0.0}_{-0.0}$ & -3194.37$^{+0.0}_{-0.0}$ & -40.09$^{+0.0}_{-0.0}$ & -5627.89$^{+0.0}_{-0.0}$ \\
c11 & -261.08$^{+0.0}_{-0.0}$ & -107.17$^{+0.0}_{-0.0}$ & -21.81$^{+0.0}_{-0.0}$ & 19.49$^{+0.0}_{-0.0}$ \\
c12 & -32.37$^{+0.0}_{-0.0}$ & -29.91$^{+0.0}_{-0.0}$ & -26.45$^{+0.0}_{-0.0}$ & -46.44$^{+0.0}_{-0.0}$ \\
c13 & -0.8781$^{+0.0082}_{-0.0082}$ & -0.814$^{+0.1572}_{-0.1546}$ & -0.9668$^{+0.0346}_{-0.0345}$ & -0.5582$^{+0.1085}_{-0.1082}$ \\
c14 & -0.0856$^{+0.0099}_{-0.0101}$ & -0.0597$^{+0.0134}_{-0.0138}$ & -0.1006$^{+0.0204}_{-0.0201}$ & -0.1831$^{+0.042}_{-0.0412}$ \\
c20 & 326.25$^{+0.0}_{-0.0}$ & 814.84$^{+0.0}_{-0.0}$ & 8.51$^{+0.0}_{-0.0}$ & 1431.78$^{+0.0}_{-0.0}$ \\
c21 & 50.12$^{+0.0}_{-0.0}$ & 21.99$^{+0.0}_{-0.0}$ & 6.95$^{+0.0}_{-0.0}$ & 1.6$^{+0.0}_{-0.0}$ \\
c22 & 4.0419$^{+0.0037}_{-0.0036}$ & 3.797$^{+0.0043}_{-0.0044}$ & 3.4251$^{+0.0058}_{-0.0055}$ & 5.9723$^{+0.0117}_{-0.0118}$ \\
c23 & 0.0544$^{+0.001}_{-0.001}$ & 0.054$^{+0.0096}_{-0.0098}$ & 0.0766$^{+0.0045}_{-0.0045}$ & 0.0708$^{+0.0141}_{-0.0142}$ \\
c30 & -41.72$^{+0.0}_{-0.0}$ & -103.24$^{+0.0}_{-0.0}$ & -0.54$^{+0.0}_{-0.0}$ & -181.59$^{+0.0}_{-0.0}$ \\
c31 & -4.24$^{+0.0}_{-0.0}$ & -1.97$^{+0.0}_{-0.0}$ & -0.81$^{+0.0}_{-0.0}$ & -0.58$^{+0.0}_{-0.0}$ \\
c32 & -0.1651$^{+0.0004}_{-0.0004}$ & -0.1585$^{+0.0005}_{-0.0005}$ & -0.1466$^{+0.0006}_{-0.0006}$ & -0.256$^{+0.0014}_{-0.0014}$ \\
c40 & 2.63$^{+0.0}_{-0.0}$ & 6.5$^{+0.0}_{-0.0}$ & -0.02$^{+0.0}_{-0.0}$ & 11.48$^{+0.0}_{-0.0}$ \\
c41 & 0.133$^{+0.0}_{-0.0}$ & 0.0651$^{+0.0}_{-0.0}$ & 0.0316$^{+0.0}_{-0.0}$ & 0.0316$^{+0.0}_{-0.0}$ \\
c50 & -0.07$^{+0.0}_{-0.0}$ & -0.1627$^{+0.0}_{-0.0}$ & 0.0$^{+0.0}_{-0.0}$ & -0.2896$^{+0.0}_{-0.0}$ \\
\hline
RMSE (dex)$^{\ast\ast}$ & 0.12 &0.07 & 0.05& 0.03\\
\hline
\end{tabular}
\tablefoot{$^{\ast} \, \rm y(X,Y) = \sum_{i=0}^{n}\sum_{j=0}^{n-i} c_{ij} \cdot X^{i} \cdot Y^{j}, \quad n=0,\dots5$, where $\rm X = [12+log(O/H)]$, $\rm Y = log_{10}{SFR}$. Fixed parameters are indicated by zero uncertainties\\
$^{\ast\ast}$RMSE represents the Root Mean Square Error of our fits per HDI.}
\end{table*}

\begin{figure*}
        \includegraphics[width=0.50\textwidth, trim={5cm 0cm 3cm 2cm},clip]{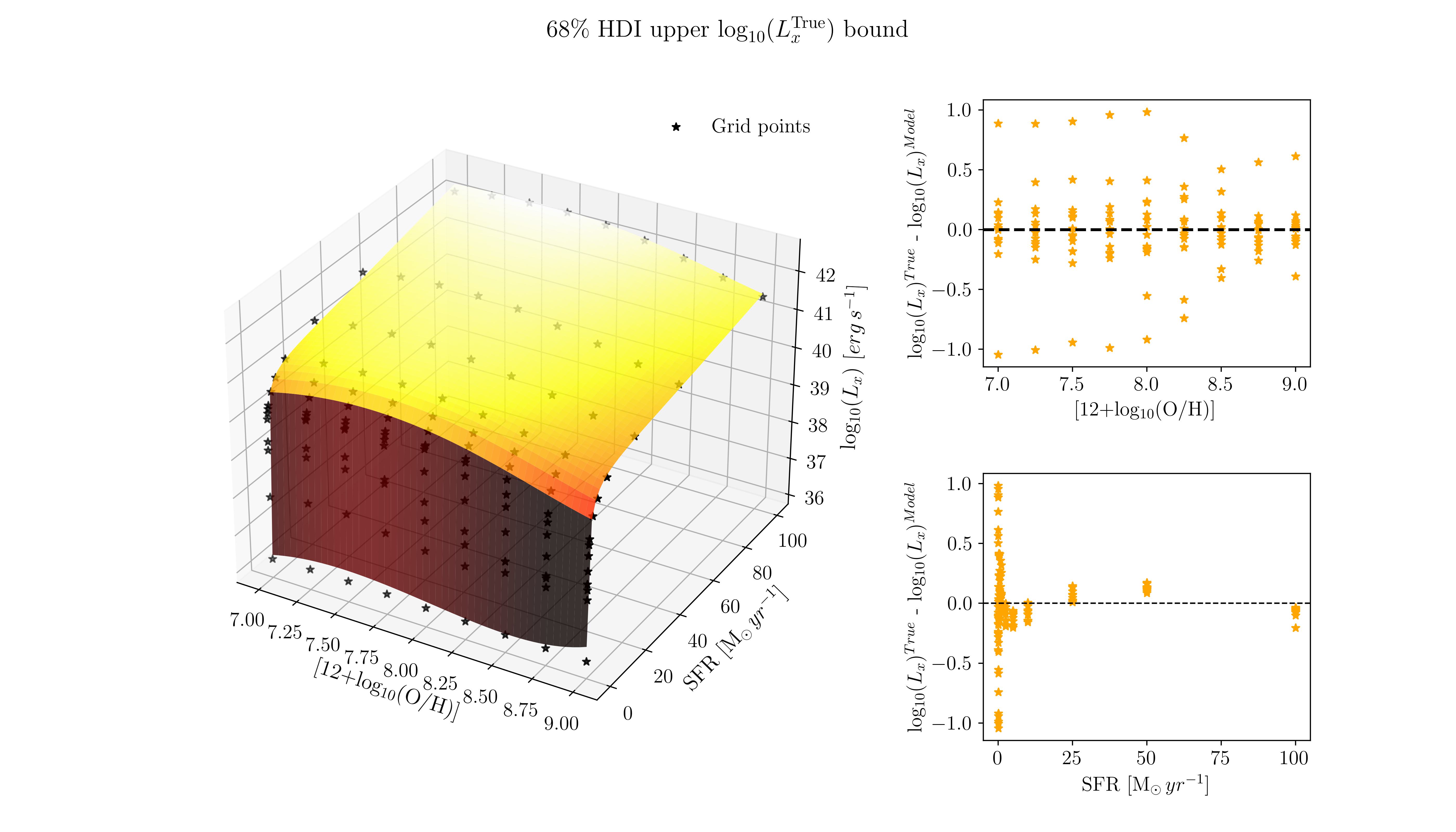}
        \includegraphics[width=0.50\textwidth, trim={5cm 0cm 3cm 2cm},clip]{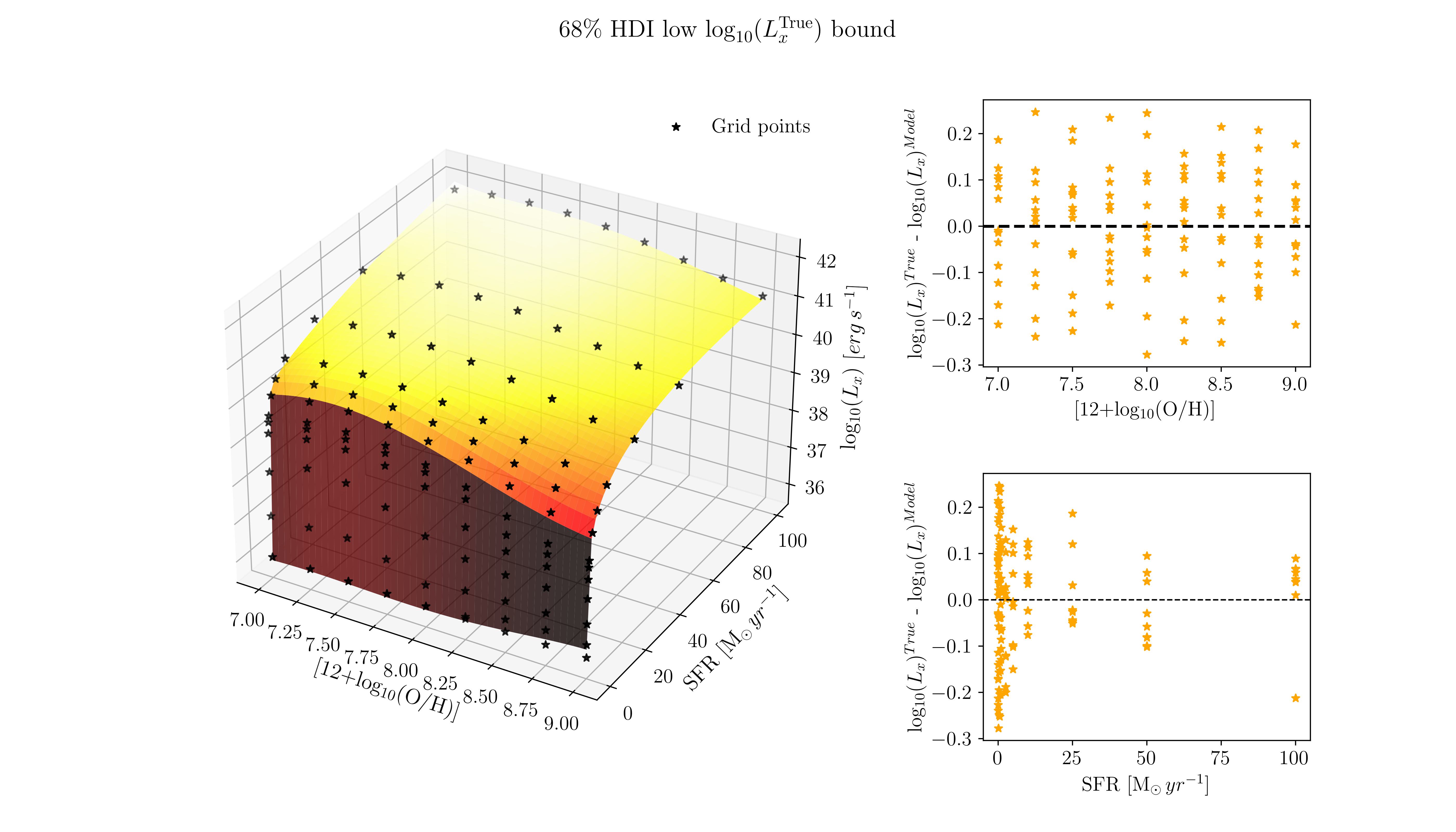}
    \caption{Same as Fig.~\ref{fig:upper_lower_surface_99} but for the 68\% HDI.}
    \label{fig:upper_surface_68}
\end{figure*}

\begin{figure*}
        \includegraphics[width=0.50\textwidth, trim={5cm 0cm 3cm 2cm},clip]{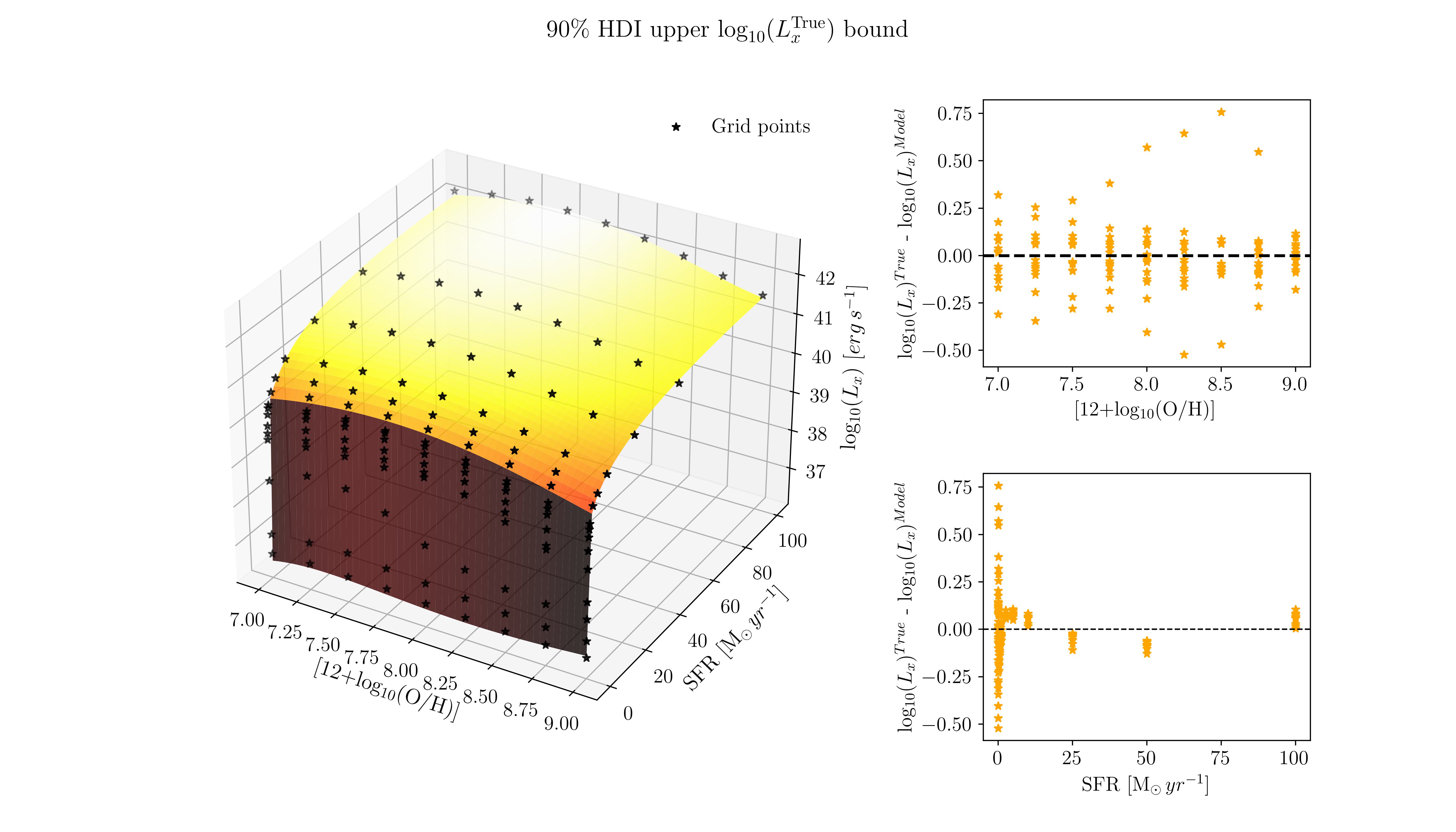}
        \includegraphics[width=0.50\textwidth, trim={5cm 0cm 3cm 2cm},clip]{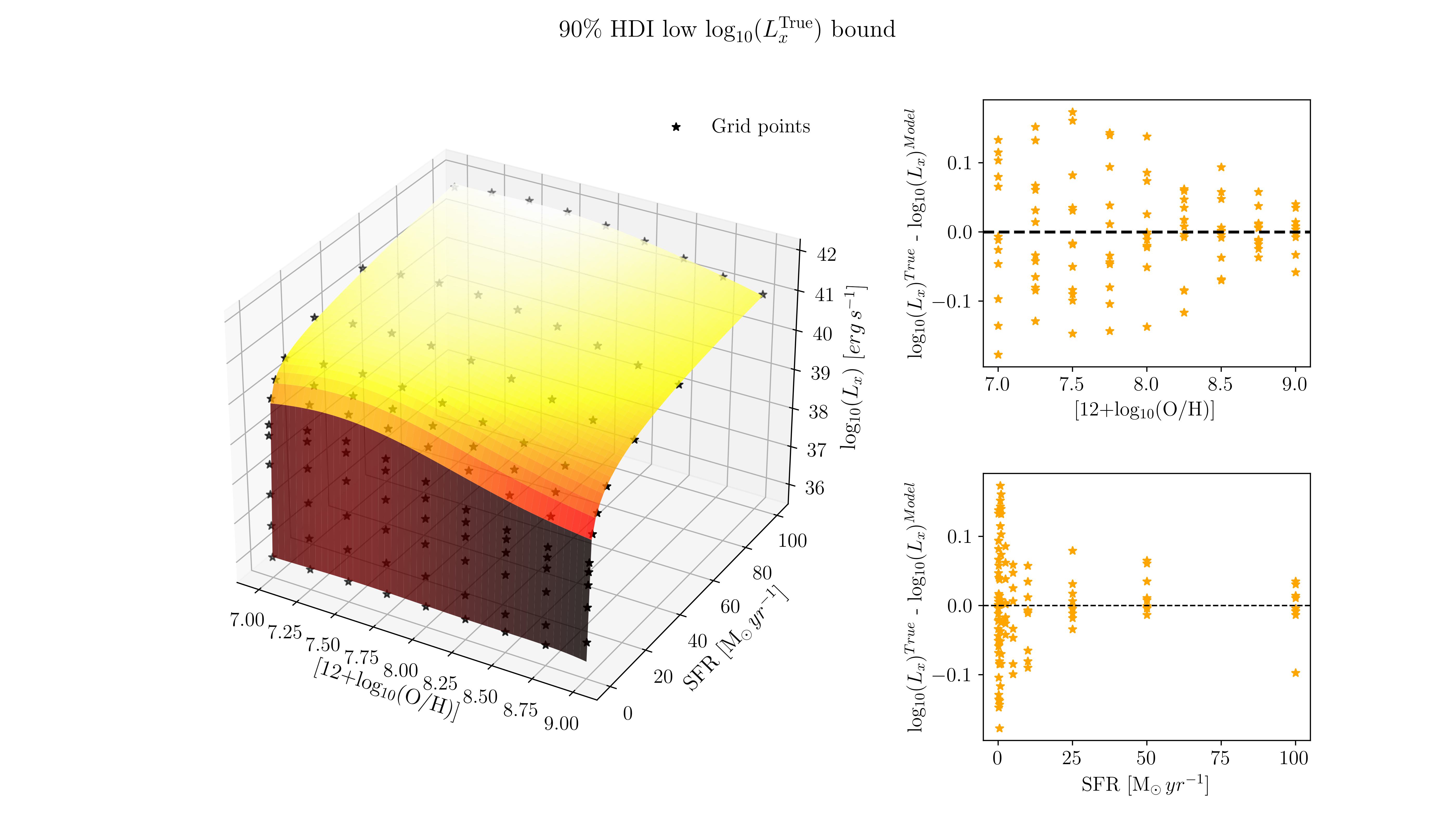}
\caption{Same as Fig.~\ref{fig:upper_lower_surface_99} but for the 90\% HDI.}
    \label{fig:upper_surface_90}
\end{figure*}

\begin{figure*}
        \includegraphics[width=0.50\textwidth, trim={5cm 0cm 3cm 2cm},clip]{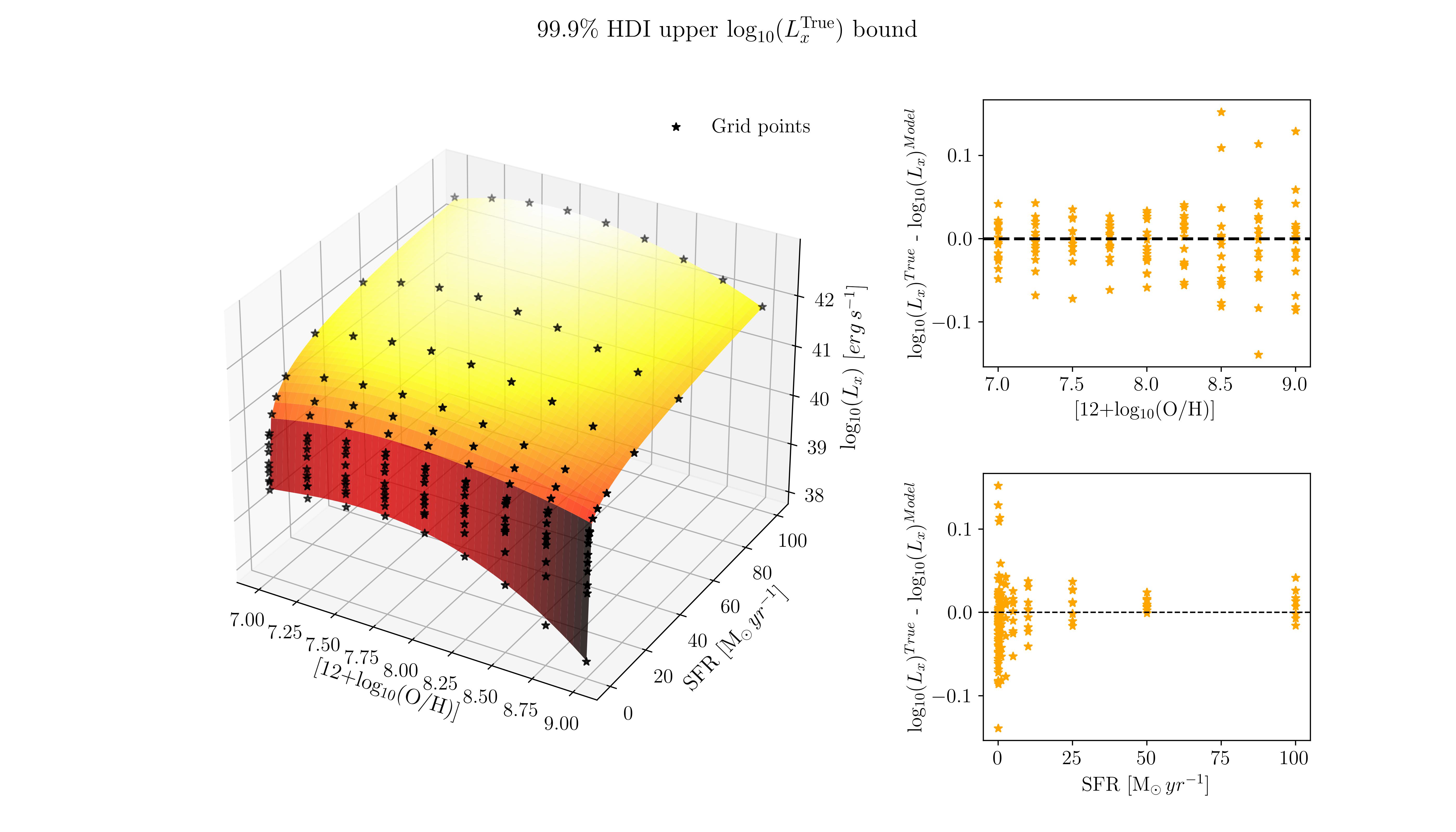}
        \includegraphics[width=0.50\textwidth, trim={5cm 0cm 3cm 2cm},clip]{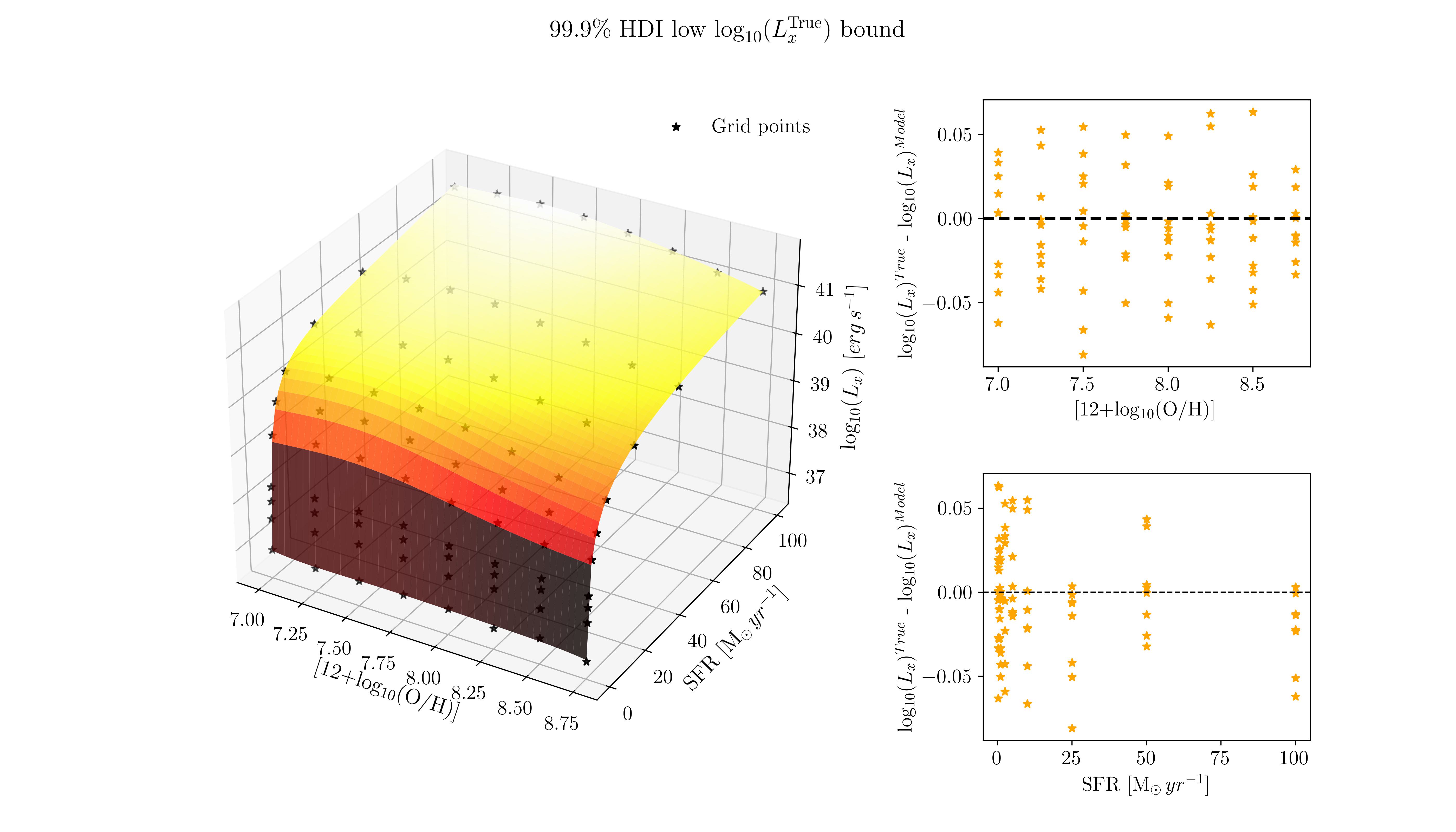}
    \caption{Same as Fig.~\ref{fig:upper_lower_surface_99} but for the 99.9\% HDI.}
    \label{fig:upper_surface_999}
\end{figure*}
%%%%%%%%%%%%%%%%%%%%%%%%%%%%%%%%%%%%%

% % Don't change these lines
% \bsp	% typesetting comment
% \label{lastpage}

\end{appendix}
\end{document}